%% file: EcoFed.tex
\documentclass[10pt,journal,compsoc]{IEEEtran}

\usepackage{cite}
\usepackage{amsmath,amssymb,amsfonts}
\usepackage{graphicx}
\usepackage{textcomp}
\usepackage[table]{xcolor}
\usepackage[ruled,linesnumbered]{algorithm2e}
\usepackage{algpseudocode}
\usepackage{commath}
\usepackage{multirow}
\usepackage{tabularx,booktabs}
\usepackage{multirow}
\usepackage{tikz}
\usepackage{caption}
\usepackage{subcaption}
\usepackage{placeins}
\usepackage{bm}
\usepackage{makecell}
\usepackage{hyperref}
\usepackage{balance}
\usepackage{amsmath,amssymb,amsfonts}
\usepackage{pifont}
\usepackage{subcaption}

\newcommand{\xmark}{\ding{55}}
\newcommand{\cmark}{\ding{51}}

\def\BibTeX{{\rm B\kern-.05em{\sc i\kern-.025em b}\kern-.08em
    T\kern-.1667em\lower.7ex\hbox{E}\kern-.125emX}}

\newcommand{\mytitle}{\texttt{EcoFed}\xspace}
\newcommand{\di}[1]{{\color{black} #1}}

\newcolumntype{P}[1]{>{\centering\arraybackslash}p{#1}}
\newcommand{\cir}[1]{\tikz[baseline]{%
    \node[anchor=base, fill=black, draw, circle, inner sep=0, minimum width=1.1em]{#1};}}
\newcommand\inv[1]{#1\raisebox{1.15ex}{$\scriptscriptstyle-\!1$}}
\RestyleAlgo{ruled}
\newcommand{\squeezeup}{\vspace{-2mm}}

\begin{document}

\title{\texttt{EcoFed}: Efficient Communication for DNN Partitioning-based Federated Learning}

\author{
    Di~Wu,
    Rehmat~Ullah,
    Philip~Rodgers,
    Peter~Kilpatrick,
    Ivor~Spence,
    and~Blesson~Varghese
    \IEEEcompsocitemizethanks{
        \IEEEcompsocthanksitem D. Wu and B. Varghese are with the School of Computer Science, University of St Andrews, UK. Corresponding author: dw217@st-andrews.ac.uk.
        \IEEEcompsocthanksitem R. Ullah is with the Cardiff School of Technologies, Cardiff Metropolitan University, UK.
        \IEEEcompsocthanksitem P. Rodgers is with Rakuten Mobile, Inc., Japan.
        \IEEEcompsocthanksitem P. Kilpatrick and I. Spence are with the School of Electronics, Electrical Engineering and Computer Science, Queen's University Belfast, UK.
    }
}


\IEEEtitleabstractindextext{
\input{Sections/abstract}
}

\maketitle
\IEEEdisplaynontitleabstractindextext
\IEEEpeerreviewmaketitle

\section{Introduction}
\label{sec:introduction}
\input{Sections/introduction}

\section{EcoFed Framework}
\label{sec:EcoFed}
\input{Sections/EcoFed}

\section{Convergence and Cost Analysis}
\label{sec:theoretical_analysis}
\input{Sections/theoretical_analysis}

\section{Evaluation}
\label{sec:evaluation}
\input{Sections/evaluation}

\section{Related Work}
\label{sec:related_work}
\input{Sections/related_work}

\section{Conclusion}
\label{sec:conclusion}
\input{Sections/conclusion}

\section{Acknowledgement}
\label{sec:acknowledgement}
\input{Sections/acknowledgement}

\bibliographystyle{IEEEtran}
\bibliography{EcoFed}

\input{Sections/Authorsbiography}

\input{Sections/Appendix_Trans}
\end{document}

%% file: Sections/abstract.tex
\begin{abstract}
Efficiently running federated learning (FL) on resource-constrained devices is challenging since they are required to train computationally intensive deep neural networks (DNN) independently. DNN partitioning-based FL (DPFL) has been proposed as one mechanism to accelerate training where the layers of a DNN (or computation) are offloaded from the device to the server. However, this creates significant communication overheads since the \di{intermediate} activation and gradient need to be transferred between the device and the server during training. While current research reduces the communication introduced by DNN partitioning using local loss-based methods, we demonstrate that these methods are ineffective in improving the overall efficiency (communication overhead and training speed) of a DPFL system. This is because they suffer from accuracy degradation and ignore the communication costs incurred when transferring the activation from the device to the server. This \di{article} proposes \mytitle~--~a communication efficient framework for DPFL systems. \mytitle eliminates the transmission of the gradient by developing pre-trained initialization of the DNN model on the device for the first time. This reduces the accuracy degradation seen in local loss-based methods.
In addition, \mytitle proposes a novel replay buffer mechanism and implements a quantization-based compression technique to reduce the transmission of the activation. 
It is experimentally demonstrated that \mytitle can reduce the communication cost by up to \di{133$\times$} and \di{accelerate} training by up to \di{21$\times$} when compared to classic FL. Compared to vanilla DPFL, \mytitle achieves a \di{16$\times$} communication reduction and \di{2.86$\times$} training \di{time speed-up}. \di{\mytitle is available from \texttt{\url{https://github.com/blessonvar/EcoFed}}}.
\end{abstract}

%% file: Sections/introduction.tex
Federated learning (FL) is a privacy-preserving machine learning paradigm that facilitates collaborative training without transferring raw data from participating devices to a server~\cite{hard2018federated,ramaswamy2019federated,leroy2019federated}.
However, running FL training on resource constrained devices is challenging since 
training deep neural networks (DNN), which is computationally expensive, is solely run on devices. This is a known bottleneck~\cite{gao2020end,wang2020towards,xu2021helios}.

\textit{\textbf{DNN partitioning-based FL}} (\textit{\textbf{DPFL}}) in which the DNN is partitioned across the device and server has been developed to surmount the challenge of running FL on resource constrained devices~\cite{wu2022fedadapt,hanaccelerating,he2020group}. In DPFL, an entire DNN model is partitioned into two parts -- a device-side model and a server-side model. The first few layers of the DNN are deployed on the device-side for training. The remaining layers are offloaded to a server that has more computational resources than the device. The computational burden on the device is alleviated as it only trains a few layers of the entire model. Consequently, the training time is reduced.

\di{Although DPFL reduces the computational burden on a device compared to FL, it incurs additional communication overheads. This is because the outputs of the activation generated by the device-side model in a forward pass and the corresponding gradients calculated during backpropagation need to be transferred between the devices and the server. Figure~\ref{figure:motivation_2} shows the computation and communication latency incurred with DPFL training for the partitioning point with minimum latency\footnote{\di{Refer to Section~\ref{sec:evaluation:setup} for the experiment configuration}}. The communication overhead in DPFL is nearly 46\% (58\% for ResNet9) of the overall training time under 5G conditions and around 86\% (90\% for ResNet9) for 3G bandwidth. In addition, this inefficiency leads to a substantial increase in: (i) the total volume of communication that is directly proportional to the data transferred across all devices and the server, and (ii) the communication frequency due to the need for transmitting activations and gradients during each forward and backward pass.
}
\input{Results/motivation}

\begin{figure*}
		\centering
		\includegraphics[width=\textwidth]{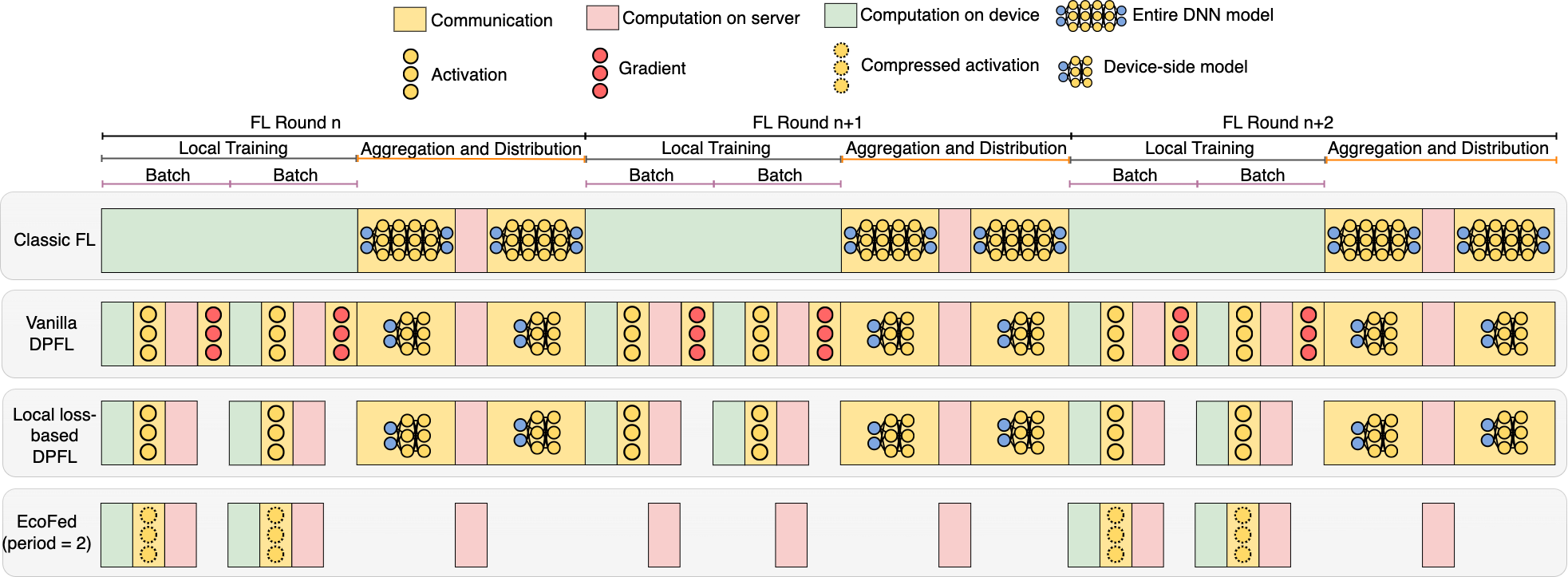}
		\caption{\di{The training pipeline of classic FL, vanilla DPFL, local loss-based DPFL and \mytitle for three rounds of training. Classic FL transfers the entire model from the devices to the server at the end of each round. Vanilla DPFL only needs to upload a partitioned device-side model at the end of each round. However, Vanilla DPFL transfers the activation and gradient for each batch sample. Local loss-based DPFL reduces the communication by half since the gradients are computed locally. \mytitle reduces communication \di{further} as it transfers the activation only periodically (for example, once in two rounds) and further compresses the size of the activations.}}
		\label{fig:comparison_dpfl}
\end{figure*}

Split Federated Learning (SFL), which we refer to as \textit{vanilla DPFL}, is the first FL work that partitions the DNN across the device and the server~\cite{thapa2022splitfed}. However, the communication overheads introduced by partitioning are not considered \di{there}. Recent DPFL methods~\cite{hanaccelerating,he2020group}, \di{which} we refer to as \textit{local loss-based DPFL}, use local loss generated by an auxiliary network to train the device-side model instead of transferring and using the gradient from the server.
Local loss-based DPFL can reduce half of the communication cost using device-side auxiliary networks since only the activation needs to be \di{sent} from the devices to the server.

Although local loss-based DPFL methods eliminate the \di{need to send the gradient}, we demonstrate that they cannot improve the efficiency of vanilla DPFL due to two significant issues (Section~\ref{sec:evaluation}).
Firstly, \textit{the volume of communication required to achieve the target accuracy is not reduced and even increases due to poor learning performance \di{as indicated} by lower final accuracy and convergence speed.} The poor learning performance of local loss-based DPFL methods is referred to as `accuracy degradation' \di{and} is caused by training using local loss instead of end-to-end training. We also demonstrate that the use of local error signals will lower the final accuracy and convergence speed (Section~\ref{sec:evaluation:acc}). As a result, the communication cost to achieve the target accuracy is similar to or even higher than vanilla DPFL (Section~\ref{sec:evaluation:comm_cost}).

\di{Secondly,} \textit{under limited network bandwidth conditions, the training time does not significantly decrease since the upload bandwidth is usually lower than the download bandwidth.} Local loss-based DPFL methods do not consider the \di{transfer of activations}, which is significant - half of the communication volume -  and has a high frequency per iteration. Therefore, when network bandwidth is limited \di{as seen in} resource constrained environments, it is not feasible to accelerate training using current local loss-based DPFL methods \di{(Section~\ref{sec:evaluation:latency_energy})}.

In this \di{article}, we propose \mytitle, a communication efficient framework for DPFL on resource-constrained devices. Figure~\ref{fig:comparison_dpfl} illustrates the training pipeline of classic FL, vanilla DPFL, local loss-based DPFL and \mytitle. \mytitle only transfers the activation periodically (for example, once every two rounds) and further reduces the size of the activation, thereby reducing the overall frequency and volume of communication. Pre-trained initialization of the device-side model is employed in \mytitle to reduce accuracy degradation caused by local loss-based methods. The frequency of transferring activations is reduced by proposing a replay buffer mechanism in which the server-side model is periodically trained by making use of cached activations instead of regularly transferring the activation from the devices to the server. Moreover, \mytitle compresses the activation using a lightweight quantization technique to further reduce the size of the data transferred and the corresponding buffer. Two DNN models and datasets are considered in our evaluations by comparing \mytitle against four baselines, including classical FL, vanilla DPFL and two state-of-the-art local loss-based DPFL methods. 
\mytitle improves the test accuracy compared to the baselines while reducing the communication volume by up to \di{16$\times$} and thus accelerates training by up to \di{2.86$\times$} compared to other DPFL methods.

The \textbf{\textit{research contributions}} of this \di{article} are:

1) Identifying the limitations of local loss-based DPFL approaches by systematically exploring the accuracy, communication size and training latency of DPFL methods on resource constrained devices.

2) Designing, developing and evaluating \mytitle, the first framework to effectively reduce communication overheads of DPFL by proposing novel approaches that optimize the forward and backward passes in DPFL.

3) Proposing novel techniques that use pre-trained initialization on the device-side to eliminate the need for transferring gradients from the server to the device without significant accuracy degradation and a replay buffer along with quantization for reducing the frequency and volume of activations transferred to the server in DPFL.

The rest of this article is organized as follows:
Section~\ref{sec:EcoFed} provides an overview and the underlying methods of the \mytitle framework.
Section~\ref{sec:theoretical_analysis} theoretically analyzes convergence and the computation and communication cost of \mytitle. 
Section~\ref{sec:evaluation} evaluates \mytitle against the baselines.
Section~\ref{sec:related_work} presents related work and
Section~\ref{sec:conclusion} concludes this \di{article}.

%% file: Results/motivation.tex
\begin{figure}[tp]
     \centering
     \begin{subfigure}[b]{0.24\textwidth}
         \centering
         \includegraphics[width=\textwidth]{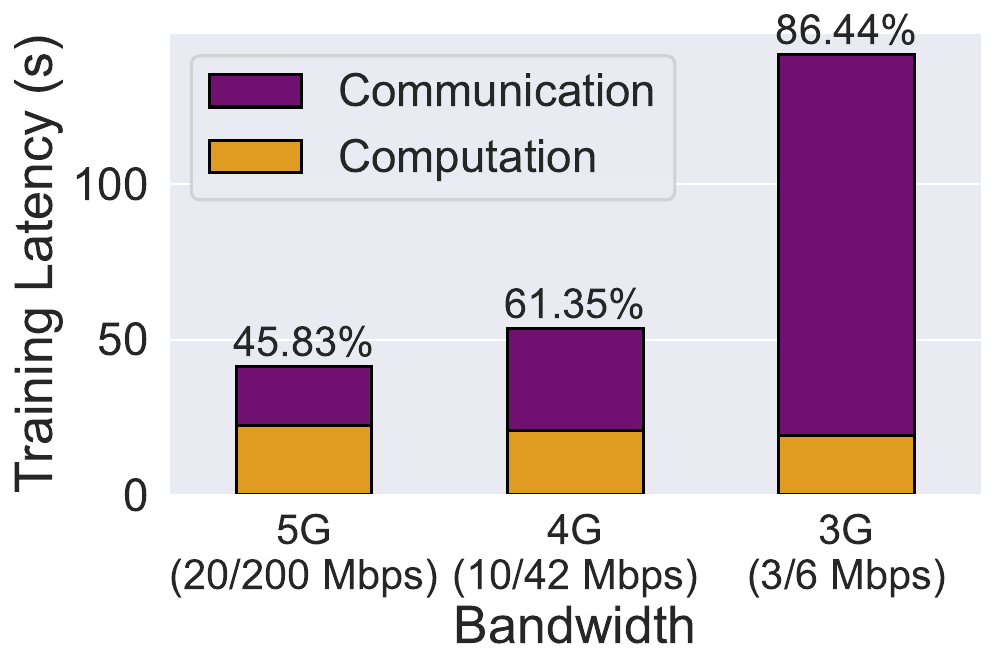}
         \caption{VGG11}
         \label{fig3_1}
     \end{subfigure}
     \hfill
     \begin{subfigure}[b]{0.24\textwidth}
         \centering
         \includegraphics[width=\textwidth]{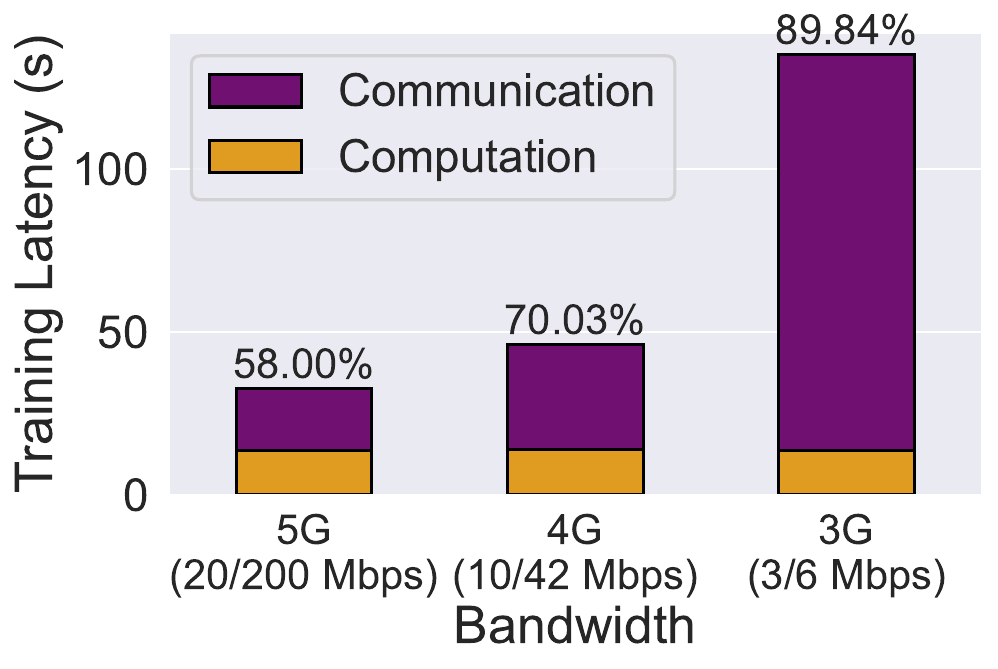}
         \caption{ResNet9}
         \label{fig3_2}
     \end{subfigure}
        \caption{\di{Computation and communication latency in DPFL training under typical (upload/download) network bandwidth. Numerical value above the bars is the percentage of communication latency.}}
        \label{figure:motivation_2}
        \squeezeup
\end{figure}

%% file: Sections/EcoFed.tex
This section firstly provides an overview of the proposed \mytitle framework (Section~\ref{sec:EcoFed:subsec:overview}). Then the underlying techniques, namely \textit{Pre-trained Initialization} (Section~\ref{sec:EcoFed:subsec:pre-trained_init}) and \textit{Replay Buffer} (Section~\ref{sec:EcoFed:subsec:replay_buffer}) are presented followed by the proposed algorithm of \mytitle (Section~\ref{sec:overallpipeline}).  

\subsection{Overview}
\label{sec:EcoFed:subsec:overview}

\begin{figure}
		\centering
		\includegraphics[width=0.48\textwidth]{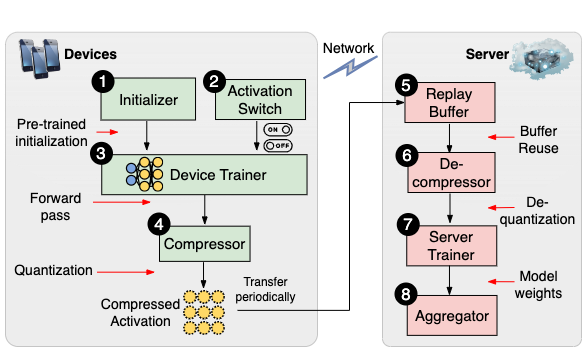}
		\caption{\di{\mytitle modules on the device and server.}}
		\label{fig:EcoFed_architecture}
\end{figure}

Figure~\ref{fig:EcoFed_architecture} provides an overview of the \di{\mytitle modules} that operate across resource constrained devices and servers (either cloud or edge servers). The underlying techniques used by the modules are discussed in the next \di{sub-sections}.

When FL training begins, the \textit{Initializer} ({\color{white}\cir{1}}) module on the server determines the training scheme (the configurations of the DNN models) and initializes the weights. The \textit{Initializer} also splits the model (to device-side and server-side models) for each participating \di{device}. The device-side models are sent to the devices and the corresponding server-side models are sent to \di{the server}.

In vanilla DPFL, after configuration, the training starts on the device-side model for each device. The \textit{Device Trainer} (the training engine on devices, {\color{white}\cir{3}}) will first generate activation outputs of the device-side model. The outputs and labels of the corresponding data samples are sent to the server. The \textit{Server Trainer} (the training engine on \di{the server}, {\color{white}\cir{7}}) trains the server-side model using the activation outputs received from the device and generates corresponding gradients. The gradients are sent back to each device to update the device-side model. The above steps are repeated for each batch sample in vanilla DPFL training.

However, in \mytitle, before generating and sending the activation outputs of the device-side model to \di{the server}, the \textit{Activation Switch} ({\color{white}\cir{2}}) will determine whether the outputs of the device-side model are required to be sent to the server or \di{if} the server can use the buffer with the cached activation to train the server-side model. If the activation outputs are required to be sent, then they will be further compressed using the quantization technique implemented by the \textit{Compressor} ({\color{white}\cir{4}}). The compressed activation and labels of the corresponding samples will then be sent to the server. On the server, the compressed data will be firstly used to update the \textit{Replay Buffer} ({\color{white}\cir{5}}) and reconstructed by the \textit{Decompressor} ({\color{white}\cir{6}}) for training the server-side models. The \textit{Server Trainer} only needs to calculate and update the gradients of the server-side models without sending the gradient back to each device for training the device-side models.

After completing the above steps, the \textit{Aggregrator} ({\color{white}\cir{8}}) will collect updated weights from each device for aggregation and for generating new global weights using the \textit{FederatedAveraging} (FedAvg) algorithm~\cite{mcmahan2017communication}.

\mytitle reduces the communication cost in vanilla DPFL by eliminating the transmission of the gradient, adjusting the communication frequency of the activation and compressing the activation data. The \textit{Initializer} uses pre-trained weights to initialize the device-side model and freezes them during training, thus eliminating the need for transmitting the gradient. In addition, \di{the} \textit{Activation Switch} periodically uploads the output activation of the device-side model to \di{the} edge servers to update the \textit{Replay Buffer}. When the edge server does not receive the activation from devices, it will continue training the server-side models using the activations that were cached in the \textit{Replay Buffer}. The \textit{Compressor} and \textit{Decompressor} modules are underpinned by quantization and dequantization techniques presented in the literature~\cite{han2015deep,choukroun2019low}. 

\subsection{Pre-trained Initialization}
\label{sec:EcoFed:subsec:pre-trained_init}

Existing local loss-based DPFL incorporates local error signals to train the device-side model using an auxiliary network~\cite{hanaccelerating,he2020group}. The auxiliary network consists of a few lightweight hidden layers, for example, fully connected layers that map the output of the device-side model to the same dimensions as the ground truth labels. Although the local error signals eliminate the need to transfer global gradients from the server to devices, this approach adversely impacts accuracy and convergence rate since the device-side model and the server-side model are decoupled and trained by different error signals, namely local and global error signals. We empirically demonstrate this in Section~\ref{sec:evaluation:acc}. A similar accuracy loss is reported in the literature for greedy layer-wise learning. In this method, the use of an auxiliary network results in accuracy degradation when compared to end-to-end training that uses global loss calculated with the entire model~\cite{belilovsky2019greedy}. As a result, the communication required to
achieve the target accuracy is effectively not reduced and even increases due to the lower accuracy and slower convergence speed of local loss-based methods.

To address the above, \mytitle adopts pre-trained initialization of the device-side model ($\bm{w}_C$). \di{For each device $k$}, the device-side model is initialized with pre-trained weights ($\bm{w}_{C,k}$), which are the partial weights of the entire model trained on a \di{pre-training} dataset (\di{e.g.,} ImageNet~\cite{deng2009imagenet}).
We empirically study the impact of local loss on generating activation outputs, which are used to train the server-side model. We identify that local loss is not a satisfactory criterion for representing a `good' activation output for training the server-side model with pre-trained initialization (Section~\ref{sec:evaluation:acc}). Therefore, during FL training, we freeze the weights of the device-side model ($\bm{w}_{C,k}$), which runs the first few layers of the model that learn general features. \di{These layers are not specific to a particular task, and therefore, the weights of the initial layers can be transferred from a pre-trained model~\cite{yosinski2014transferable,huh2016makes}}.

\textbf{Benefits of pre-trained initialization:} There are four benefits to pre-trained initialization and freezing weights of the device-side model $\bm{w}_{C,k}$ in DPFL. They are: (i) Communication during training is reduced since the need for the devices to receive gradients from the edge server is eliminated; (ii) \di{Accuracy} loss caused by using local error signals is reduced; (iii) \di{The} computational workload on resource constrained devices is reduced since the gradients of $\bm{w}_{C,k}$ do not need to be calculated and updated on the devices; (iv) \di{The} device-side model ($\bm{w}_{C,k}$) and the respective activation outputs are not significantly changed during each training round, making it possible to use cached activations from a buffer to train server-side models.

\subsection{Replay Buffer}
\label{sec:EcoFed:subsec:replay_buffer}
\mytitle introduces a \textit{Replay Buffer} on the edge server, i.e., the server uses the activation cached in the buffer, which is obtained from a previous training round, to train the server-side model in a given round. 
\mytitle caches and updates the buffer periodically. 
If the transmission of activations is switched off in a given round, then \mytitle will reuse the cached activations. Therefore, there are two modes of training, namely \mytitle with and without the buffer.

\textbf{Periodic transfer:} The \textit{Activation Switch} controls the transfer frequency of the activation in the forward pass from the device-side to the server-side model. The frequency is controlled by the interval between successive activation transfers, which is denoted as $\rho$. Consider, for example, $\rho = 2$. In this case, the activation outputs of the devices will only be transferred every second round and the buffer is cached with the activations when it is sent. During the other rounds, the edge server will use the cached activations to train the server-side models. It is worth noting that the activations ($\bm{a}_k^t$) for each round will change due to different participating clients, data batches and data augmentations. \di{Therefore}, the buffer on the edge servers needs to be periodically updated during the training. Thus, $\rho$ is a hyper-parameter that will affect the model performance and communication cost of \mytitle, which is further considered in Section~\ref{sec:evaluation:acc}.   

\textbf{Reducing the buffer size:} A potential issue that must be mitigated is that a large buffer may be required if large or many activations are transferred. The maximum size of the buffer required will be the size of all activations transferred from all devices. However, it is practical to establish a maximum buffer size and implement periodic updates to the buffer to efficiently manage its storage capacity. In addition, \mytitle stores the compressed activation outputs ($\bm{z}$) instead of the original activations ($\bm{a}_k^t$). The activation is compressed by the \textit{Compressor} module of \mytitle using a lightweight 8-bit linear quantization technique~\cite{wu2020easyquant} denoted as function $Q(.)$. The output ($\bm{a}_{k}^t$) is quantized from 32 bits to 8 bits before sending it to the server. The compressed activation ($\bm{z}_k^t$) is then cached in the \textit{Replay Buffer}.

\begin{algorithm}[tp]
\caption{Partitioning-based training in \mytitle}
\label{algorithm1}
\textbf{Input:} Pre-trained $\bm{w}_C^*$ and data $\{D_k\}_{k=1}^{K}$\\
\textbf{Output:} $\bm{w}^*$\\

 \For{each device $k \in K$ in parallel}{
    Download $\bm{w}_C^*$ to device $k$;\\
    Initialize and freeze $\bm{w}_{C,k}^*$;{\color{teal} \small{\null\hfill//Pre-trained initialization}}\\
 }

\For{each round (each device and corresponding worker $k \in K$ in parallel)}{
        Initialize $\bm{w}_{S,k}$ on the server;\\
            \For{each round $t \in T$}{
                \If {Activation switch is on ($t \mod \rho == 0$)}{
                    Forward on $\bm{w}_{C,k}^*$ to get $\bm{a}_{k}^{t}$;\\
                    $\bm{z}_k^t \gets Q(\bm{a}_{k}^{t})$;
                    {\color{teal} \small{\null\hfill//Quantization}}\\
                    Send $\bm{z}_k^t$ to the server;\\
                    update buffer with $\bm{z}_k^t$;
                    }
                \Else{
                    $\bm{z}_k^t \gets $ buffer; 
                    {\color{teal}\small{\null\hfill//Replay Buffer}}\\
                    }
                $\bm{\hat{a}}_{k}^{t} \gets \inv{Q}(\bm{z}_k^t)$; {\color{teal} \raggedright \small{\null\hfill//Dequantization}}\\
                Forward on $\bm{w}_{S,k}^{t}$;\\
                Calculate loss $\ell_{S,k}^{t}$ and gradients $\nabla \ell_{S,k}^{t}(\bm{w}_{S,k}^{t})$;\\
                $\bm{w}_{S,k}^{t} \gets \bm{w}_{S,k}^{t} - \eta \nabla \ell_{S,k}^{t}(\bm{w}_{S,k}^{t})$;\\
                }
        $\bm{w}_{S,k} \gets \bm{w}_{S,k}^{t}$;\\
        $\bm{w}_{S} \gets \Sigma_{k=1}^{K} \frac{|D_k|}{|D|}\bm{w}_{S,k}$; {\color{teal}\small{\null\hfill//FedAvg}}\\
}
$\bm{w}_S^* \gets \bm{w}_S$;\\
$\bm{w}^* \gets \{\bm{w}_C^*, \bm{w}_S^*\}$; 
{\color{teal}\small{\null\hfill//Concatenation of $\bm{w}_C^*$ and $ \bm{w}_S^*$}}\\
\Return{$\bm{w}^*$}
\end{algorithm}

\subsection{Proposed Algorithm}
\label{sec:overallpipeline}

Algorithm~\ref{algorithm1} shows the steps of partitioning-based training in \mytitle by employing pre-trained initialization and a replay buffer with compression using quantization.

We first present the notation used. The collection of $K$ devices is denoted as $\{k\}_{k=1}^{K}$. Each device generates its own data $D_k$. The data from all devices is denoted as $D \colon \{D_k\}_{k=1}^K$. The number of samples in $D_k$ is denoted as $\vert D_k \vert$ and the total number of samples is $\vert D \vert$. Let $\bm{w}$ be the entire model, which will be partitioned as the device-side model ($\bm{w}_C$) and server-side model ($\bm{w}_S$). $\bm{w}_{C,k}$ and $\bm{w}_{S,k}$ are the models of the $k^{th}$ device. The superscript $t$ is used to represent a training round $t$ in a total of $T$ rounds. $\bm{a}_k$ is the intermediate activation generated by $\bm{w}_{C,k}$. $\ell_{S,k}(.)$ is the loss function of the server-side model of the $k^{th}$ device.

\mytitle first prepares the pre-trained weights of the device-side model ($\bm{w}_C^*$). The weights of $\bm{w}_C^*$ will be frozen during training to eliminate the transfer of the gradient from the server-side to device-side model $\bm{w}_C$ (Lines 4 and 5). Then, the server-side model of each device will be trained independently with $K$ parallel workers. 

For each round $t$, if activation transfer is switched on (i.e. $t \mod \rho == 0$), then the output activation of $\bm{w}_{C,k}^*$ is generated, denoted as $\bm{a}_k^t$, and compressed to 8 bits, denoted as $\bm{z}_t^k$. The compressed activation is uploaded to the edge server and is used to update the corresponding buffer (Lines 10-15). Alternatively, the edge server retrieves $\bm{z}_t^k$ directly from the buffer (Line 17). Using $\bm{z}_t^k$, \mytitle will firstly dequantize $\bm{z}_t^k$ and provide the output $\bm{\hat{a}}_{k}^{t}$ to the server-side model for the training of $\bm{w}_{S,k}$ (Lines 19-22). 

Once a training round is completed by an edge server, it will send each $\bm{w}_{S,k}$ to the cloud. The cloud aggregates a new global server-side model (Lines 24-26). Finally, after $T$ rounds of training, the cloud obtains the optimal server-side model ($\bm{w}_S^*$) and concatenates it with $\bm{w}_C^*$ into a complete model $\bm{w}*$ (Lines 28-29).

%% file: Sections/theoretical_analysis.tex
In this section, we firstly analyze \di{\mytitle} convergence (Section~\ref{sec:ActionFed:subsec:convergence_analysis}). Then the computation and communication costs of \mytitle on the device-side are compared against classic FL, vanilla DPFL and local loss-based DPFL (Section~\ref{sec:ActionFed:subsec:cost_analysis}).

\subsection{Convergence Analysis}
\label{sec:ActionFed:subsec:convergence_analysis}
We follow the convergence analysis presented in the literature~\cite{hanaccelerating, belilovsky2020decoupled,wang2022fedlite} to analyze the convergence of Algorithm~\ref{algorithm1}.
We assume the following:

\textbf{Assumption 1} -- \textit{The server-side objective functions are L-smooth, i.e., $\lVert \nabla F_S(\bm{u}) - \nabla F_S(\bm{v}) \rVert \leq L \lVert \bm{u} - \bm{v} \rVert, \forall \bm{u}, \forall \bm{v}$}.

\textbf{Assumption 2} -- \textit{The squared norm of the stochastic gradient has an upper bound for the server-side object function, i.e., $\lVert \nabla F_S (\bm{a}_k^t;\bm{w}_{S,k}^t) \rVert^2 \leq G, \forall k, \forall t$}.

\textbf{Assumption 3} -- \textit{The learning rate $\eta_t$ satisfies $\sum_t \eta_t = \infty$ and $\sum_t \eta_t^2 < \infty$}.

In each global round $t$, the output activation of the $k_{th}$ device-side model is equal to $\bm{a}_{k}^t = \bm{w}_{C,k}^t(\bm{x})$. We assume $\bm{a}_{k}^t$ follows a probability distribution of $p_{k}^t(\bm{a})$, which is determined by $\bm{w}_C^t$ and $D_k$. In \mytitle, the device-side model $\bm{w}_{C,k}^t$ is initialized with pre-trained weights and frozen during the training, thus fixed as $\bm{w}_{C,k}^*$. \mytitle uses the quantization function $Q(.)$ and dequantization function $\inv{Q}(.)$ along with the \textit{Replay Buffer} to train the server-side models, denoted as $\bm{\hat{a}}_{k}^t \triangleq \inv{Q}(Q(\bm{a}_{k}^t))$. We define the probability distributions of the \textit{Replay Buffer} as $q_{k}^t(\bm{a})$. 

To analyze the impact of  \textit{Replay Buffer} and quantization on the convergence, we further define two types of errors: buffer and quantization errors. Buffer error is defined as the distance of gradients between the original distribution $p_{k}^t(\bm{a})$ and buffer distribution $q_{k}^t(\bm{a})$, denoted as $\delta_{k} \triangleq \int  \left\lVert \nabla \ell ( a_k^t;\bm{w}_S) \right\lVert  \left\lVert (q_{k}^t(\bm{a}) - p_{k}^t(\bm{a})) 
\right\lVert da$. The $\ell (.)$ is the loss function (e.g. cross entropy loss for classification). 
Quantization error is defined as $\int \left\lVert \nabla \ell ( \hat{a_k^t};\bm{w}_S) - \nabla \ell ( a_k^t;\bm{w}_S) \right\lVert  \left\lVert q_{k}^t(a)  \right\lVert da$, which is the sum of gradient errors over $q_{k}^t(\bm{a})$ due to quantization. We include a further assumption specific to \mytitle as follows:

\textbf{Assumption 4} -- \textit{The buffer error and the quantization error have upper bounds, i.e., $\delta_k^t \leq H_1 $ and $ \varepsilon_k^t \leq H_2, \forall k, \forall t$.}

\textbf{Convergence of device-side model:} 
Since $\bm{w}_{C}$ is fixed during training, its convergence is not considered. 

\textbf{Convergence of server-side model:} Let $\frac{1}{\Gamma_T} \triangleq \sum_{t=0}^{T-1} \eta_t$. \textit{Following on from Assumption~1, Assumption~2 and  Assumption~4, Algorithm~\ref{algorithm1} ensures the following: } 

\begin{align*}
\frac{1}{\Gamma_T} \sum_{t=0}^{T-1} \eta_t \mathbb{E} [ \lVert \nabla F_S(\bm{w}_S^{t}) \rVert^2 ] 
\leq 
\frac{4 ( F_S(\bm{w}_S^{0}) - F_S(\bm{w}_S^{*}) )}{3 \Gamma_T} \\
+ 
\frac{1}{\Gamma_T} \sum_{t=0}^{T-1} \left( \eta_t (\sqrt{G} + 1) (H_1+H_2) + 
\frac{L}{2} \eta_t^2 G\right)
\tag{1}
\label{eq1}
\end{align*}
where $\nabla F_S(\bm{w}_S^{t}) \triangleq \frac{1}{K} \sum_{k=1}^K \nabla F_{S,k}(\bm{w}_S^{t})$. $F_{S,k}(.)$ is the objective function of the $k^{th}$ server-side model. $\bm{w}_S^{0}$ is the initial server-side weights and $\bm{w}_S^{*}$ is the optimal server-side weights. The detailed proof is provided in Appendix A.

Based on Assumption 3, it is noted that with increasing $T$, the right-hand side of Equation~\ref{eq1} convergences to zero. Thus, Equation~\ref{eq1} guarantees that the proposed algorithm of \mytitle converges to a stationary point with increasing $T$.

\textbf{Differences between the convergence of local loss-based DPFL and \mytitle:} 
Equation~\ref{eq1} is similar to the convergence analysis presented in the literature~\cite{hanaccelerating}. The server-side model converges as follows: 
\begin{align*}
\frac{1}{\Gamma_T} \sum_{t=0}^{T-1} \eta_t \mathbb{E} [ \lVert \nabla F_S(\bm{w}_S^{t}) \rVert^2 ] 
\leq 
\frac{4 ( F_S(\bm{w}_S^{0}) - F_S(\bm{w}_S^{*}) )}{3 \Gamma_T} \\
+ 
G \frac{1}{\Gamma_T} \sum_{t=0}^{T-1} \left( \eta_t \frac{1}{K} \sum_{k=1}^K \left( d_{k}^t \right)
+ 
\frac{L}{2} \eta_t^2 \right)
\tag{3}
\label{eq3}
\end{align*}
where $d_{k}^t \triangleq \int \lVert p_{k}^t(\bm{a}) - p_{k}^*(\bm{a}) \rVert d\bm{a} $ which is defined as the distance between the probability distribution of activation $\bm{a}_k^t$ and $\bm{a}_k^*$. It is worth noting that $p_{k}^t(\bm{a})$ keeps changing during FL training since $\bm{w}_{C,k}^t$ is updated by the local error signals. Therefore, in local loss-based DPFL, the changing distance ($d_{k}^t$) of the probability distribution of $\bm{a}_k^t$ caused by updating $\bm{w}_{C,k}^t$ affects the convergence of the server-side model as shown in Equation~\ref{eq3}. However, in \mytitle, $d_{k}^t = 0$ since $\bm{w}_{C,k}^t$ is fixed during training. In addition, the \textit{Replay Buffer} and quantization error affects convergence behaviour, which is $\delta_{k}^t $ and $\varepsilon_k^t$ (bounded by $H_1$ and $H_2$) as shown in Equation~\ref{eq1}. Different $\rho$ values and quantization techniques used by the \textit{Compressor} module will determine the values of $H_1$ and $H_2$, thus affecting the convergence of \mytitle.

\subsection{Cost Analysis}
\label{sec:ActionFed:subsec:cost_analysis}
Table~\ref{table:theoretical_comparison} compares the computation and communication \di{costs} of \mytitle against classic FL, vanilla DPFL and local loss-based DPFL. We \di{use} $\vert . \vert$ \di{to denote} either the computation or communication workload of a given model or an activation. We distinguish two modes in \mytitle, namely \mytitle without buffer and \mytitle with buffer. In classic FL, the entire model ($\bm{w}$) is computed on each device ($\vert \bm{w} \vert$). At the end of each round, $\bm{w}$ is uploaded to the cloud and then the newly aggregated $\bm{w}$ is downloaded to the device ($2 \vert \bm{w} \vert$). For vanilla DPFL, each device \di{trains} \di{only} the device-side model, $\bm{w}_C$ ($\vert \bm{w}_{C} \vert$) and $\bm{w}_C$ is transferred at the end of each round ($2 \vert \bm{w}_{C} \vert$). \di{The} activation and gradient of each data sample will be communicated with an overhead of $2 \vert D_k \vert \vert \bm{a}_k \vert$. For local loss-based DPFL, each device also \di{needs} to train $\bm{w}_C$ ($\vert \bm{w}_{C} \vert$). In addition, $\bm{w}_C$ is uploaded and downloaded at the end of each round ($2 \vert \bm{w}_{C} \vert$) but only the activation is transferred during \di{each training round} ($\vert D_k \vert \vert \bm{a}_k \vert$). 

However, for \mytitle without using the Replay Buffer (indicated as \mytitle w/o buffer; when the buffer is used indicated as \mytitle w buffer), the device only computes the forward pass on $\bm{w}_C$, ($ \frac{1}{2} \vert \bm{w}_{C} \vert$). In addition, the communication overhead is reduced to $\vert D_k \vert \vert \bm{z}_k \vert$ where $\bm{z}_k$ is the compressed activation~\footnote{\di{\label{exception}There is one exception - in the first round of training, devices need to download $\bm{w}_{C}$. Therefore, the communication cost for the first round is $\vert \bm{w}_{C} \vert + \vert D_k \vert \vert \bm{z}_k \vert$.}}. There are no computation or communication costs during training when using the buffer.

\begin{table}
\begin{center}
\caption{Computation and communication costs on the device for each round.}
\begin{tabular}{ P{3cm}  P{1.5cm} P{2.5cm} }
\Xhline{2\arrayrulewidth}
\textbf{Methods}  &\textbf{Computation}  &\textbf{Communication}\\
\hline
 FL  &$\vert\bm{w}\vert$  &$2 \vert\bm{w}\vert$\\

Vanilla DPFL  &$\vert\bm{w}_{C}\vert$  &$2 \vert\bm{w}_{C}\vert + 2 \vert D_k \vert \vert\bm{a}_k\vert $\\

Local loss-based DPFL  &$\vert\bm{w}_{C}\vert$  &$2\vert\bm{w}_{C}\vert + \vert D_k \vert \vert\bm{a}_k\vert$\\

\hline
\textbf{\mytitle w/o buffer}  &$\frac{1}{2}\vert\bm{w}_{C}\vert$   &$\vert D_k \vert\vert\bm{z}_k\vert$~\ref{exception}\\

\textbf{\mytitle w buffer}  &$0$  &$0$\\
\Xhline{2\arrayrulewidth}
\end{tabular}
\label{table:theoretical_comparison}
\squeezeup
\end{center}
\end{table}

%% file: Sections/evaluation.tex
In this section, we evaluate \mytitle against four baselines. The test environment, the evaluation setup and the results obtained from the experimental studies are considered. The results highlight that \mytitle improves accuracy compared to state-of-the-art methods and eliminates accuracy degradation caused by local error signals in local loss-based DPFL. In addition, \mytitle significantly reduces communication costs and accelerates training.

\subsection{Test Environment}
\label{sec:evaluation:testbed}

\textbf{Datasets and Models:} Two image classification datasets, namely CIFAR-10~\cite{krizhevsky2009learning} and CIFAR-100~\cite{krizhevsky2009learning} are used \footnote{We do not report learning performance on MNIST~\cite{lecun1998gradient} and FMNIST~\cite{xiao2017fashion} as they are simple datasets and do not highlight the differences between the methods we evaluate.} We follow a similar method reported in the literature~\cite{mcmahan2017communication} to partition data across devices. For an independent and identically distributed (I.I.D.) setting, the training set is initially divided into 500 shards for CIFAR-10 and 5000 shards for CIFAR-100. We randomly assign 5 shards and 50 shards to each device for CIFAR-10 and CIFAR-100, respectively. In the non-I.I.D. setting, we first sort the dataset based on their labels, dividing it into 500 shards for CIFAR-10 and 5000 shards for CIFAR-100. Then, we randomly allocate 5 shards and 50 shards to each device for CIFAR-10 and CIFAR-100, respectively. Consequently, each device will have training samples of up to half of all available classes. The test dataset is available on the server and will be used to test model performance after each round of training. 

We train two popular convolutional neural networks, namely VGG11~\cite{simonyan2014very} (plain convolutional neural network) and ResNet9~\cite{he2016deep} (residual convolutional neural network). \di{In terms of the DNN partitioning point (PP) and auxiliary networks used on devices, we follow the same configuration used in the local loss-based DPFL literature~\cite{hanaccelerating}, which corresponds to PP2 detailed in Appendix B.3. The architectures of VGG11, ResNet9 and the auxiliary network along with the device-side and server-side model partitions are shown in Table~\ref{table:model_archi}}.

\textbf{FL Training hyper-parameters:}
At the beginning of each FL round, the server randomly selects 20 devices out of 100 devices (sampling ratio of 0.2) for participating in the current training round. The standard FedAvg~\cite{mcmahan2017communication} aggregation method is used by the \textit{Aggregator} for all approaches. We adopt the same data augmentation and learning schedules for all methods for fair comparisons (horizontal flip with a probability of 0.5 and the Stochastic Gradient Descent (SGD) optimizer with a learning rate of 0.01). We set a total of 500 rounds for training on all datasets.

\textbf{Testbed:}
To \di{evaluate} system performance (i.e. communication cost and training latency), we build a prototype with an edge server and five resource constrained devices. The edge server has a 2.5GHz Intel i7 8-core CPU and 16GB RAM. Five Raspberry Pi 4 Model B single-board computers with 1.5GHz quad-core ARM Cortex-A53 CPU are used as resource constrained devices. All devices and \di{server run PyTorch as the training engine}. The server and devices are connected \di{via} a router. We use Linux \texttt{tc} commands to emulate different network \di{(upload/download) bandwidths: 5G (20/200 Mbps), 4G (10/42 Mbps) and 3G (3/6 Mbps) as indicated in Appendix B.1}.

We also employ a simulation-based \di{testbed comprising} a 2GHz AMD EPYC 7713P 64-Core CPU with 252GB RAM and two Nvidia A6000 GPUs. This \di{testbed enables us to rapidly carry out the} evaluation of learning performance.

\begin{table}
	\centering
	\caption{Models for evaluation. Convolution layers denoted as C followed by the no. of filters; filter size is $3 \times 3$ for all convolution layers except for downsampling convolution \di{which is} $1 \times 1$; Max Pooling layer is MP; Fully Connected layer is FC; and Residual Block is RB including two convolution layers, a max pooling layer and downsampling convolution layers; number following is no. of output channels.}
	\begin{tabular}{ P{1.5cm} P{2.5cm} P{3.5cm} }
        \Xhline{2\arrayrulewidth}
        \textbf{Model}       &\textbf{Device}    &\textbf{Server} \\
     \hline
        VGG11        &C64-MP-C128-MP &C256-C256-MP-C512-C512-MP-C512-C512-FC4096-FC4096-FC10\\
        \hline
        ResNet9     &C64-MP-C128-MP &RB256-RB512-RB512-FC10 \\
        \hline
        Auxiliary Network    &FC10 &N/A \\
    \Xhline{2\arrayrulewidth}
    \end{tabular}
	\label{table:model_archi}
\end{table}

\input{Results/convergence_curves}
\input{Tables/acc}

\subsection{Evaluation Setup}
\label{sec:evaluation:setup} 

\textbf{Baselines:} We consider four baselines, namely classic FL, SFL (vanilla DPFL) and two state-of-the-art local loss-based DPFL methods:
(1) \textbf{Classic FL} trains the entire model locally \di{on-device}.  
(2) \textbf{SFL (vanilla DPFL)} is the first DPFL approach. The model is partitioned into device-side and server-side models, sent to the devices and the server for training, respectively. The activation and gradient are transferred between devices and the server for each data batch. This method does not optimize communication. 
(3) \textbf{Local generated loss (LGL)}~\cite{hanaccelerating} introduces a locally generated loss on devices to calculate gradients for training the device-side model, thus reducing the need for sending gradients from the server.
(4) \textbf{FedGKT}~\cite{he2020group} incorporates local loss on \di{devices but also uses} probabilistic predictions, called soft labels~\cite{hinton2015distilling}, of the server-side models. \di{These} are periodically transferred to the devices and vice versa. The soft labels distil the training of device-side and server-side models\footnote{The implementation of FedGKT for our experiments is a variant that we developed based on vanilla DPFL in order to compare it to other baselines when using the same aggregation \di{algorithm (FedAvg)}.}.
  
\textbf{EcoFed} is implemented with the following configuration. We adopt the pre-trained weights of the device-side model from the respective pre-trained model trained on the ImageNet dataset~\cite{deng2009imagenet} and freeze them during the training. $\rho = 2$ for the \textit{Replay Buffer} (the buffer will be updated every two rounds). In addition, a linear 8-bit quantization implementation is adopted for the \textit{Compressor}. 

\textbf{Evaluation Metrics:}
Two sets of metrics are used:

1) \textbf{Metrics on learning performance}: (i) \textit{Accuracy} is evaluated on the global test data. We record the highest test accuracy achieved during entire rounds of training for each baseline. The results are an average of three independent runs with different random seeds.
  
2) \textbf{Metrics on system performance}: (i) \textit{Communication cost} is the measured communication overhead for one round. The communication cost versus test accuracy curve is presented to evaluate the efficiency of communication for achieving a target accuracy; (ii) \textit{Training latency} is the wall-clock time of one training round for each baseline given a network bandwidth.

\subsection{Accuracy} 
\label{sec:evaluation:acc}
The test accuracy curves of the five methods, including \mytitle and the four baseline methods, on the two datasets (both I.I.D. and Non-I.I.D. setting) using the two DNN models are shown in Figure~\ref{figure:convergence_curves}. In addition, the highest test accuracy is reported in Table~\ref{table:acc}. The results show that \mytitle usually outperforms all baselines across all datasets and the two model architectures (except for FL on I.I.D. CIFAR-10 for ResNet9). In detail, \mytitle achieves up to a 4.63\% increase in accuracy on Non-I.I.D. CIFAR-100 for VGG11 compared to SFL (4.31\% compared to FL). In addition, \mytitle significantly improves the accuracy by up to 6.22\% on Non-I.I.D CIFAR-100 for VGG11 compared to LGL and by up to 18.08\% on Non-I.I.D CIFAR-10 for VGG11 compared to FedGKT, respectively. 

SFL offloads parts of the training computation to the server but fundamentally shares the same algorithm as FL. As a result, their accuracy and learning curves are similar. LGL and FedGKT introduce local error signals to reduce communication. However, although there is a reduction in communication, there is a significant loss in accuracy and slower convergence compared to FL and SFL. \mytitle, on the other hand, does not have an accuracy degradation seen in LGL and FedGKT, while also achieving a more substantial reduction in communication costs compared to local loss-based DPFL methods (Section~\ref{sec:evaluation:comm_cost}).

\subsubsection{Impact of pre-trained initialization and freezing weights on the device-side model}
 
To obtain a better understanding of accuracy achieved in Figure~\ref{figure:convergence_curves} and Table~\ref{table:acc}, we further investigate accuracy degradation in local loss-based DPFL methods and the effects of using pre-trained initialization and freezing weights on $\bm{w}_c$ in \mytitle. We answer the following three questions\footnote{\di{We show the results for only CIFAR-10 due to the limitation of space;} the same conclusions were observed for CIFAR-100.}:

\textbf{Can local error signals generated on the device degrade accuracy?} From Table~\ref{table:acc} and Figure~\ref{figure:convergence_curves}, it is noted that local loss-based DPFL methods suffer a higher accuracy loss than the other methods. Even when compared to FL and SFL, LGL and FedGKT consistently achieve lower accuracy. In addition, local loss-based DPFL has a slower convergence rate as demonstrated in Figure~\ref{figure:convergence_curves} resulting from separately optimizing the device-side model and the server-side model by inconsistent gradient signals.

\input{Tables/fedgkt}
We also note that FedGKT has a relatively lower accuracy compared to LGL. In FedGKT, knowledge distillation is carried out both on device-side and server-side models. We demonstrate that the distillation of device-side soft labels on server-side training has a negative impact on accuracy. Table~\ref{table:fedgkt} shows the accuracy results of bidirectional FedGKT and unidirectional FedGKT (only the server-side soft labels are used for distilling the device-side model). We observe a significant accuracy improvement when removing distillation from device-side soft labels when training the server-side model, shown as unidirectional FedGKT. This further demonstrates that local training results in accuracy degradation of the server-side model.

\input{Tables/pretrained_init}

\textbf{Can pre-trained initialization on the device-side model also improve accuracy for other baselines?}
We further investigate the impact of pre-trained initialization of the device-side model for FL, SFL and local loss-based DPFL methods. Table~\ref{table:pretrained_init} shows the highest test accuracy achieved by each method when we apply pre-trained initialization for the device-side model. The results indicate that pre-trained initialization can significantly improve the accuracy of FL and SFL, specifically in the Non-I.I.D. setting. This finding has also been observed in recent research~\cite{chen2022pre,nguyen2022begin}.

Surprisingly, the results demonstrate that pre-trained initialization can also mitigate accuracy degradation caused by local error signals in the local loss-based DPFL methods, which to the best of our knowledge, has never been reported before. When all methods adopt pre-training initialization, \mytitle still outperforms all local loss-based DPFL methods. Compared to FL and SFL, \mytitle achieves competitive accuracy performance (less than 1\% loss) with considerable communication reduction by up to 114$\times$ (Section~\ref{sec:evaluation:comm_cost}).

\input{Tables/lgl_frozon}
\textbf{Does training device-side models with local loss in the context of pre-trained initialization improve the accuracy of the server-side model?} 
In LGL and FedGKT, the device-side model is trained by local error signals to avoid \di{receiving the gradients}. However, \di{for} pre-trained initialization, we investigate whether device-side training by local loss can improve the training of the server-side model after applying pre-trained initialization on the device-side model. 

To this end, the test accuracy of device-side model ($\bm{w}_c$) and server-side model ($\bm{w}_s$) of LGL are recorded on CIFAR-10. We compare the test accuracy of LGL (with pre-trained initialization) with trainable $\bm{w}_c$ (device-side model) and frozen $\bm{w}_c^*$. Table~\ref{table:lgl_frozon} shows that the trainable $\bm{w}_c$ has significantly higher test accuracy than the frozen $\bm{w}_c^*$ across I.I.D. and non-I.I.D. settings. \di{However, it is surprising that the server-side model ($\bm{w}_s$)}, when the device-side model is frozen (denoted as $\bm{w}_s$ under $\bm{w}_c^*$), achieves higher accuracy compared to the server-side model with a trainable device-side model (denoted as $\bm{w}_s$ under $\bm{w}_c$).

The results suggest that training the device-side model can indeed significantly improve the accuracy of the device-side model, but it does not necessarily improve the accuracy of the server-side model and may even degrade the accuracy of the server-side model. However, the accuracy of the server-side model is what we ultimately aim to achieve.
It is thus inferred that local training on the device-side model is not required for improving the accuracy performance of server-side model. We conjecture that the local error signals generated by the local auxiliary network are not optimal for the training performance of server-side models. Given the above observations, \mytitle freezes the device-side weights when pre-trained initialization is adopted.

\subsubsection{Impact of $\rho$ and quantization on accuracy} 
We investigate the accuracy performance of \mytitle under different hyper-parameter settings. $\rho$ controls the update frequency of the \textit{Replay Buffer} and quantization is adopted to reduce the size of transferred data and consequently, the memory required by the cached buffer. The impact of $\rho$ and quantization on test accuracy for CIFAR-10 is shown in Figure~\ref{fig:period_quantization}. The accuracy gradually decreases as $\rho$ increases since the update frequency is reduced. Accuracy is less sensitive to quantization since (near) similar accuracy is achieved with or without quantization for the same $\rho$ value. It is worth noting that when $\rho = 1$, \mytitle updates the cached buffer every round. \di{To achieve higher accuracy, the buffer must be updated more frequently (smaller $\rho$).}

\di{
\textbf{Dynamic control strategy for $\rho$:} We explore dynamic control of $\rho$ as accuracy is sensitive to $\rho$. We use a heuristic-based strategy and monitor accuracy improvement every 40 rounds. The accuracy growth rate, denoted by $\Delta_{acc}$, is calculated. The value of $\rho$ for the next 40 rounds is set based on the predefined heuristic rules shown in Table~\ref{table:dynamic_rho}. Figure~\ref{fig:dynamic_rho} presents the highest accuracy and the corresponding change of $\rho$ during training. In general, $\rho$ gradually decreases as training accuracy approaches a plateau in the later stages. We calculate the average value of $\rho$ with dynamic control, which is $\rho = 3.5$ throughout the entire training process. In addition, it is observed that the highest accuracy using dynamic control is also obtained between fixed periods with $\rho = 2$ and $\rho = 4$.
\begin{table}
\begin{center}
\di{
\caption{\di{Heuristic rules for dynamic $\rho$.}}
\begin{tabular}{ P{1cm} P{1.3cm} P{1.3cm} P{1.3cm} P{1.5cm}}
\Xhline{2\arrayrulewidth}
Period & $\Delta_{acc} \in (10^{-1}, \infty)$ & $\Delta_{acc} \in (10^{-2}, 10^{-1}]$ & $\Delta_{acc} \in (10^{-3}, 10^{-2}]$ & $\Delta_{acc} \in (-\infty, 10^{3}]$ \\
\hline
$\rho$  & 8  & 4 & 2 & 1\\
\Xhline{2\arrayrulewidth}
\end{tabular}
\label{table:dynamic_rho}
}
\end{center}
\end{table}
}
\input{Results/dynamic_rho}
\input{Results/period_quantization}

\subsubsection{Impact of pre-training dataset}
\di{
\input{Tables/pretraining_datasets}

Pre-trained weights obtained using large-scale datasets (e.g., ImageNet) for the device-side model can be downloaded from open-source repositories, such as Hugging Face\footnote{https://huggingface.co/} and PyTorch Model Zoo\footnote{https://pytorch.org/serve/model\_zoo.html}. This approach saves substantial pre-training time when using large datasets. We have also explored the impact on accuracy when training from scratch on a small pre-training dataset, e.g., Tiny-ImageNet~\cite{le2015tiny}. The training overhead for this is minimal compared to FL training since the server usually has more resources than devices. In addition, given the challenges of collecting user data for pre-training, we have carried out pre-training on synthetic data, such as CIFAR-5m~\cite{nakkiran2020deep} and SIP-17~\cite{zhu2023towards}. Please refer to Appendix B.2 for a description of pre-training on these datasets. Table~\ref{table:pretraining_datasets} shows the accuracy results on CIFAR-10 when using pre-trained weights on four different types of pre-training datasets. In general, \mytitle demonstrates robust generalization across various pre-training datasets. It achieves a higher level of accuracy on natural datasets and maintains similar accuracy, even on small-scale datasets like Tiny-ImageNet. \mytitle exhibits a small decrease in accuracy compared to SFL (up to 0.96\%) for CIFAR-5m. However, it still outperforms local loss-based methods such as LGL (by up to 3.95\%). When dealing with the more challenging SIP-17 dataset with domain shifts, \mytitle still achieves high performance with a small accuracy loss (up to 1.83\%) compared to SFL. In addition, it outperforms local loss-based methods, such as LGL (up to 1.79\%).
}
\subsection{Communication Cost}
\label{sec:evaluation:comm_cost}

\textbf{Communication cost for one training round:} The communication overhead of \mytitle and the four baseline methods are compared for one training round on the CIFAR-10 dataset \di{as shown in Table~\ref{table:comm}}\footnote{For evaluation of communication cost and training latency of one training round, we only report the results on the CIFAR-10 dataset as system performance is independent of datasets.}. 

Compared to classic FL, other DPFL methods have a smaller communication overhead; for example, the communication cost is reduced by 8.27$\times$ (SFL) and 15.55$\times$ (LGL and FedGKT) when training VGG11. The reason is that DPFL methods only need to transfer the device-side model between the devices and the server, which is usually smaller than the entire model. \mytitle (with buffer) achieves a further reduction in communication cost of \di{66.62$\times$} on VGG11 and \di{18.18$\times$} on ResNet9. When using the buffer, \mytitle fundamentally eliminates the need for communication. In terms of the overall communication costs for all rounds of FL, \mytitle ($\rho = 2$) can reduce communication by \di{133.25$\times$} on VGG11 and \di{36.36$\times$} on ResNet9 compared to classic FL.

Compared to the other DPFL methods, \mytitle also reduces the communication cost significantly. \mytitle without buffer reduces the communication cost by \di{8.05$\times$, 4.29$\times$ and 4.29$\times$} on both VGG11 and ResNet9. LGL and FedGKT can reduce the communication cost by half when compared to SFL as they only require the activations to be transferred from devices to the server. However, due to the cached buffer technique in \mytitle, the average communication per round is further reduced by a factor of $\rho$. For instance, in our experiments with $\rho = 2$, the average communication per round is \di{0.0385 GB}, which is \di{16.1$\times$, 8.57$\times$ and 8.57$\times$} lower than SFL, LGL and FedGKT, respectively.

\textbf{Communication cost vs test accuracy:} The cumulative communication cost for all training rounds on the two datasets is considered. The communication cost after each training round is recorded. Figure~\ref{figure:commvsacc} highlights the communication cost incurred to achieve a target test accuracy.

For instance, given a test accuracy target of 80\% on I.I.D. CIFAR-10 for VGG11, FL, SFL, LGL and FedGKT will require 729GB, 90GB, 71GB and 145GB of data transfers, respectively. Similarly, for a 50\% test accuracy target on I.I.D. CIFAR-100 the communication costs for VGG11 are 1121GB, 139GB, 110GB and 156GB. However, \mytitle will only require 4GB on I.I.D. CIFAR-10 and 7GB on I.I.D. CIFAR-100 to achieve the target test accuracy.
For ResNet9, FL, SFL, and LGL will require 169GB, 78GB, and 67GB of data to be transferred (150GB, 65GB, 64GB and 100GB on I.I.D. CIFAR-100), respectively. FedGKT fails to achieve 80\% test accuracy on I.I.D. CIFAR-10 and requires 100GB on I.I.D. CIFAR-100 for 50\% test accuracy. In contrast, \mytitle has a low volume of data transfer \di{(4GB on I.I.D. CIFAR-10 and 3GB on I.I.D. CIFAR-100)} to achieve the target test accuracy. 

For the Non-I.I.D. setting on both datasets, \mytitle has higher communication efficiency to reach a target test accuracy when compared to all baselines. It is worth noting that although the communication of LGL and FedGKT is reduced by half per round (since gradient transfers are eliminated) compared to SFL, the volume of communication required to achieve the same level of accuracy does not significantly decrease, and in some cases, such as ResNet9 on CIFAR-100 increases. This is because of accuracy degradation and slower convergence using local error signals.

\input{Tables/comm}
\input{Results/latency}
\input{Results/latency_pps}

\mytitle reduces communication cost by \di{133.25$\times$} on VGG11 (\di{36.36$\times$} on ResNet9) compared to classic FL. Compared to SFL (vanilla DPFL), \mytitle improves communication by up to \di{16.1$\times$}. In short, classic FL, SFL (vanilla DPFL) and local loss-based DPFL incur significant communication costs. In contrast, \mytitle has significantly lower communication costs and higher communication efficiency.

\input{Results/comm_vs_acc}

\subsection{Training Latency}
\label{sec:evaluation:latency_energy}

\di{\mytitle improves the communication efficiency and eliminates device-side gradient computation resulting in a speed-up of the overall training latency.
The average training time for one round of \mytitle is compared against the baseline methods to quantify this benefit.} Figure~\ref{figure:latency_vgg11} and Figure~\ref{figure:latency_resnet9} highlight that compared to classic FL, \mytitle achieves a \di{9.33$\times$} and \di{10.06$\times$} speed-up for VGG11 and ResNet9 without using buffer (under 5G conditions), respectively. In addition, compared to SFL (vanilla DPFL), \mytitle achieves a speed-up of \di{1.47$\times$} and \di{1.9$\times$} on VGG11 and ResNet9 without using the buffer \di{(for 5G)}. Compared to local loss-based DPFL methods (LGL and FedGKT) with communication optimization, there is an improvement of training latency of about \di{1.38$\times$} and \di{1.91$\times$} on VGG11 and \di{1.83$\times$} and \di{2.62$\times$} on ResNet9 without using the buffer. 

We further consider the impact of network bandwidth, which will be a bottleneck for devices that operate in environments that have limited network capabilities (e.g. mobile phones with 4G or 3G signal). Therefore, we evaluated \mytitle and each baseline under different network bandwidth conditions. When the network bandwidth is limited to 4G (10 Mbps and 42 Mbps for upload and download) and 3G (3 Mbps and 6 Mbps for upload and download), the training latency of FL, SFL, LGL and FedGKT are high due to the increase in the communication costs for transferring the model and intermediate activation and gradients. It is worth noting that local loss-based methods (i.e. LGL and FedGKT) have a similar training latency compared to non-optimized vanilla DPFL (i.e. SFL) since they still incur large communication costs due to transferring activations. In addition, the upload bandwidth for sending activations is typically much lower than the download bandwidth used for sending gradients, and it has not been considered in local loss-based DPFL. This highlights the importance of reducing communication costs when transferring activations.

In contrast, \mytitle has only a small increase in training latency, resulting in a \di{21.08$\times$} and \di{2.38$\times$} speed-up on VGG11 and \di{11.26$\times$} and \di{2.86$\times$} speed-up on ResNet9 in a 3G network, compared to FL and SFL, respectively. In addition, when \mytitle employs the buffer for training (half of the entire rounds with $\rho = 2$), the training latency is fundamentally independent of network bandwidths. \di{This validates the low bandwidth requirements of \mytitle}.

\textbf{Latency under different partitioning points:}
Variants of vanilla DPFL dynamically adjust the PP to minimize training latency~\cite{deng2022low,he2023accelerated,wu2022fedadapt}. We investigated the impact on training latency for different PPs in \mytitle. We considered four PPs (PP1, PP2, PP3 and PP4); refer to the complete configuration in Appendix B.3. We compared \mytitle with SFL for different PPs using 5G network bandwidth. We report the overall latency and its individual components of computation and communication. Please refer to Appendix B.4 for more results on accuracy performance.

Figure~\ref{figure:latency_pps} illustrates the latency of SFL and \mytitle under different PPs. As the partition point is placed deeper in the model, the overall training latency gradually increases since more computations are performed on the device. In addition, the communication latency becomes smaller as activations and gradients in the later layers are smaller in size. Overall, \mytitle achieves significant acceleration across all partition points without using the buffer. It results in speed-ups of 1.31$\times$, 1.47$\times$, 2.03$\times$, and 2.83$\times$ for VGG11 (and 1.8$\times$, 1.9$\times$, 2.6$\times$, and 3.36$\times$ for ResNet9) at PP1, PP2, PP3, and PP4, respectively. When the partition point is at the initial layers of the model (PP1 and PP2), \mytitle minimizes latency by reducing network communication latency. When the partition point is at the later layers of the model (PP3 and PP4), \mytitle reduces latency by eliminating gradient computation on the device. The lowest overall latency is achieved for PP1, where \mytitle achieves 1.8$\times$ speed-up compared to SFL. When the buffer is used there is no computation on the device. In this case, the training latency of \mytitle is further reduced and gradually decreases when the PP is set at later layers since less computation is offloaded to the server. In summary, \mytitle can further accelerate training latency for different partitioning points.

%% file: Results/convergence_curves.tex
\begin{figure*}
     \centering
     \begin{subfigure}[b]{0.24\textwidth}
         \centering
         \includegraphics[width=\textwidth]{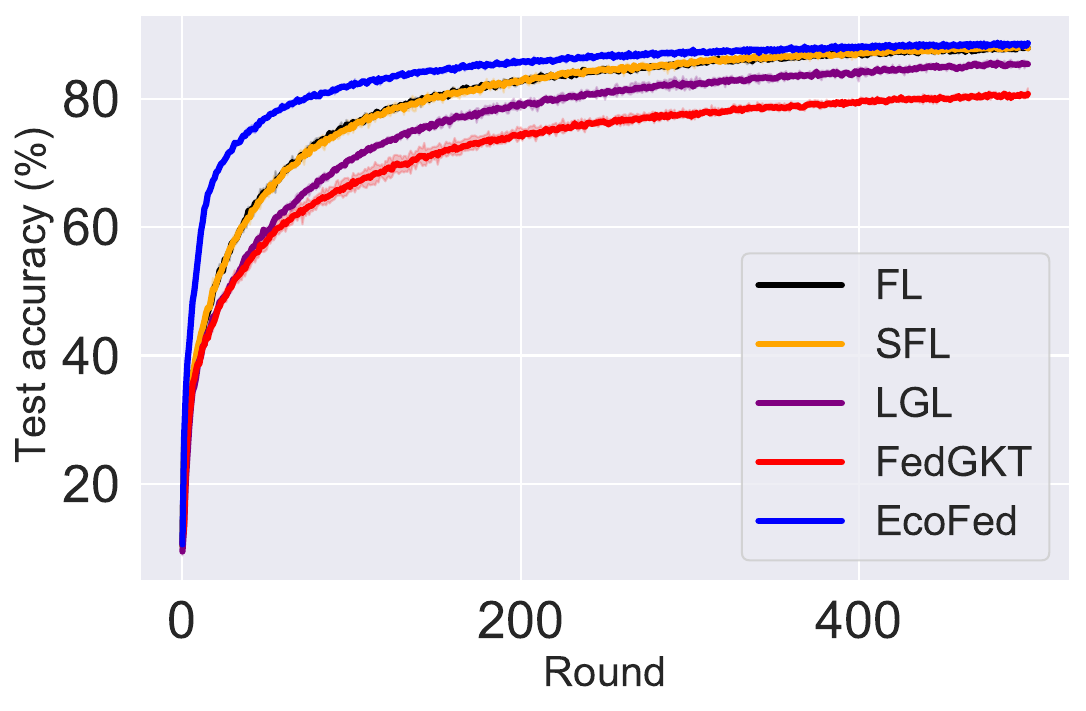}
         \caption{VGG11 on I.I.D. CIFAR-10}
         \label{}
     \end{subfigure}
     \hfill
     \begin{subfigure}[b]{0.24\textwidth}
         \centering
         \includegraphics[width=\textwidth]
         {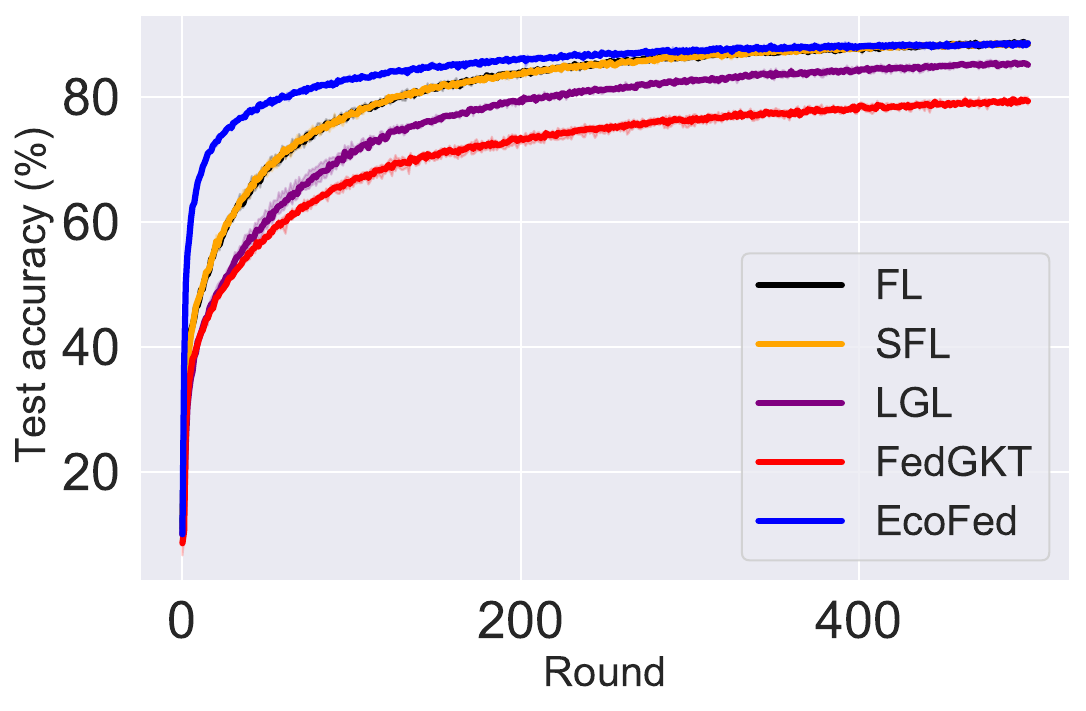}
         \caption{ResNet9 on I.I.D. CIFAR-10}
         \label{}
     \end{subfigure}
     \hfill
     \begin{subfigure}[b]{0.24\textwidth}
         \centering
         \includegraphics[width=\textwidth]
         {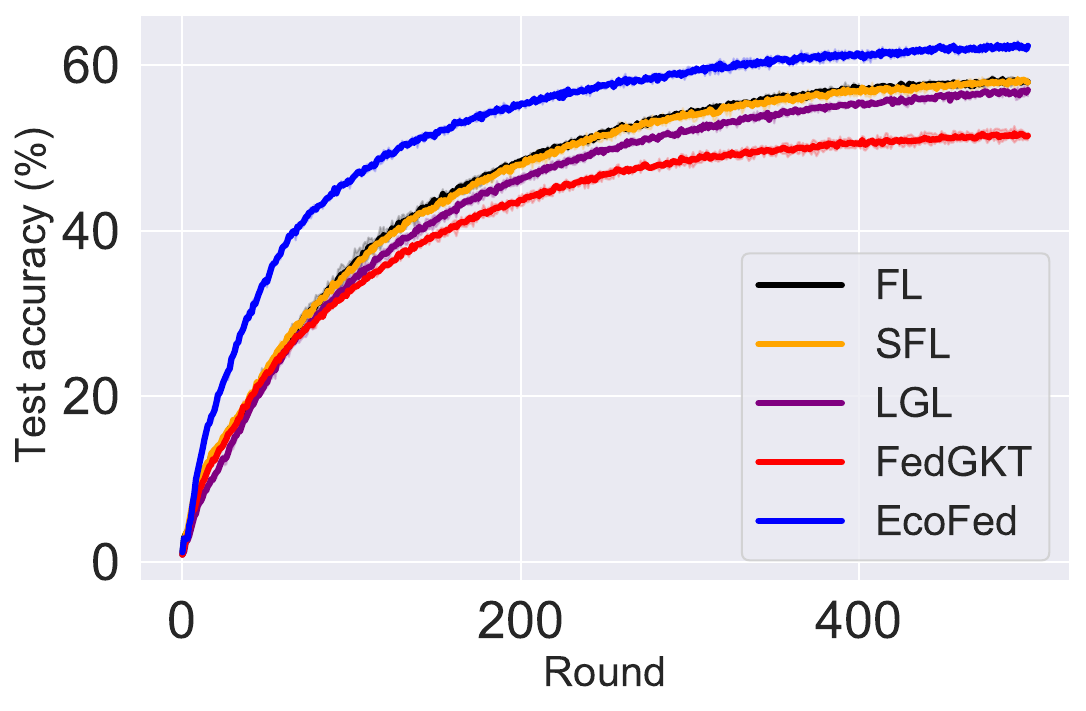}
         \caption{VGG11 on I.I.D. CIFAR-100}
         \label{}
     \end{subfigure}
     \hfill
     \begin{subfigure}[b]{0.244\textwidth}
         \centering
         \includegraphics[width=\textwidth]
         {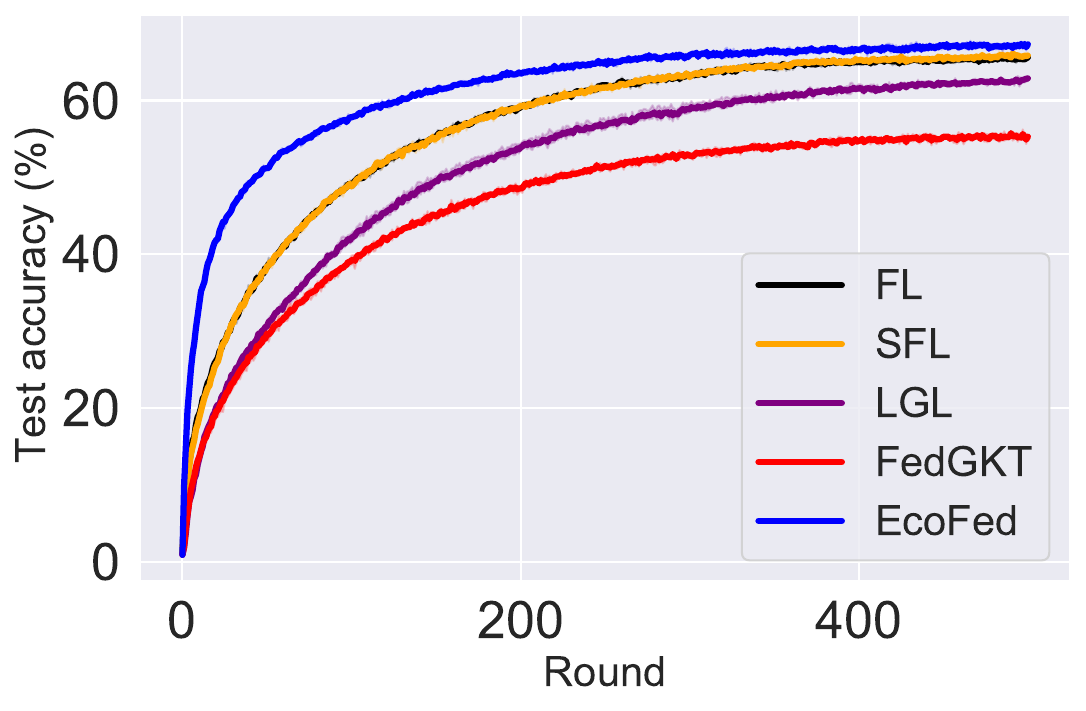}
         \caption{ResNet9 on I.I.D. CIFAR-100}
         \label{}
     \end{subfigure}
     \hfill
     \begin{subfigure}[b]{0.24\textwidth}
         \centering
         \includegraphics[width=\textwidth]
         {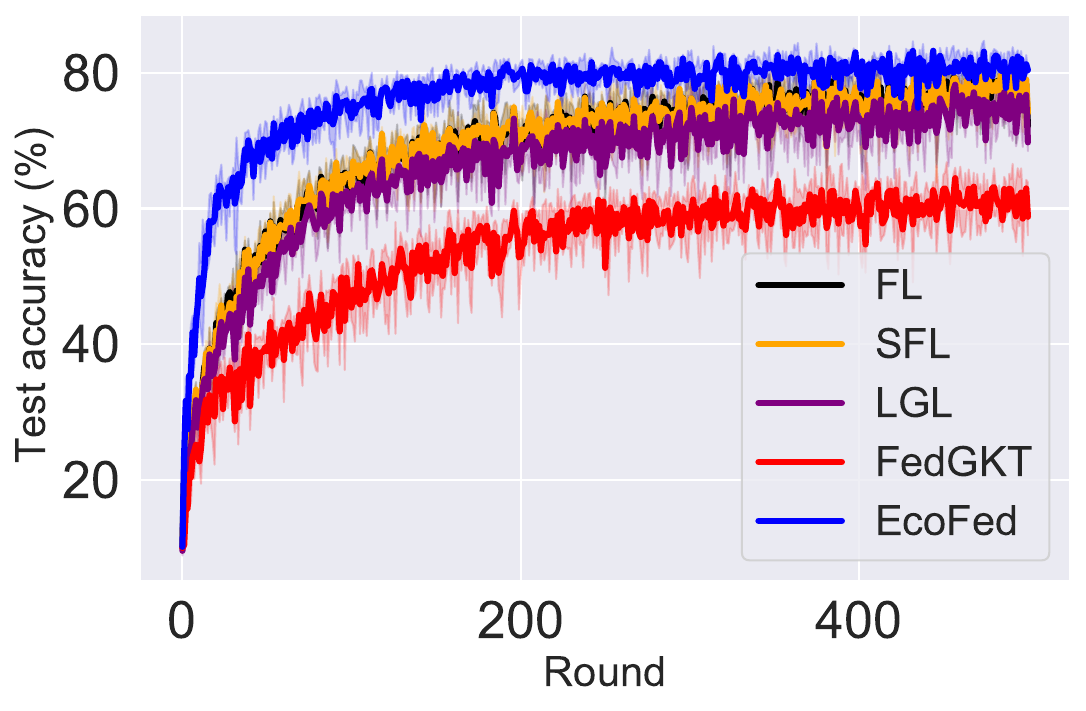}
         \caption{\centering VGG11 on Non-I.I.D. CIFAR-10}
         \label{}
     \end{subfigure}
     \hfill
     \begin{subfigure}[b]{0.24\textwidth}
         \centering
         \includegraphics[width=\textwidth]
         {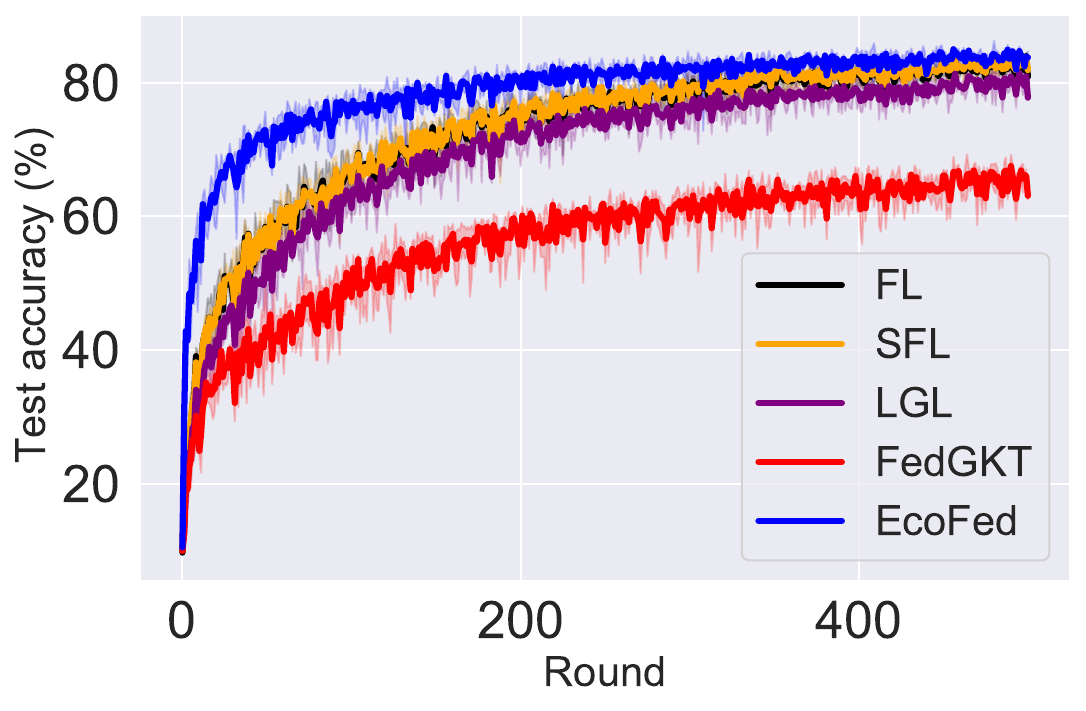}
         \caption{\centering ResNet9 on Non-I.I.D. CIFAR-10}
         \label{}
     \end{subfigure}
     \hfill
     \begin{subfigure}[b]{0.24\textwidth}
         \centering
         \includegraphics[width=\textwidth]
         {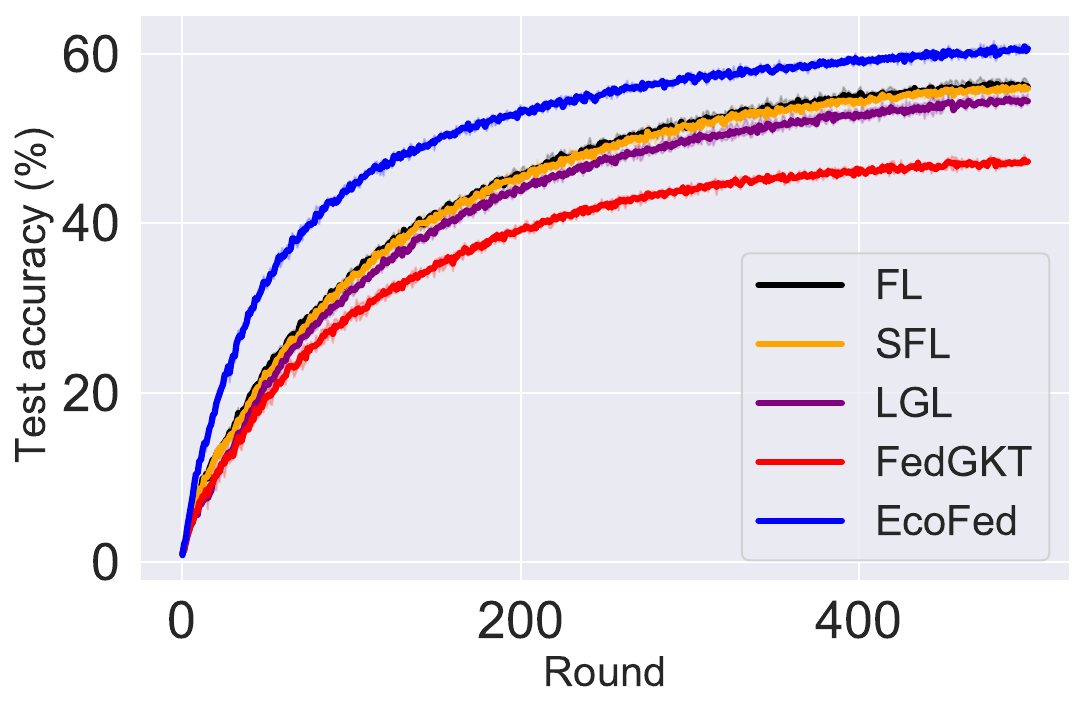}
         \caption{\centering VGG11 on Non-I.I.D. \space CIFAR-100}
         \label{}
     \end{subfigure}
     \hfill
     \begin{subfigure}[b]{0.24\textwidth}
         \centering
         \includegraphics[width=\textwidth]
         {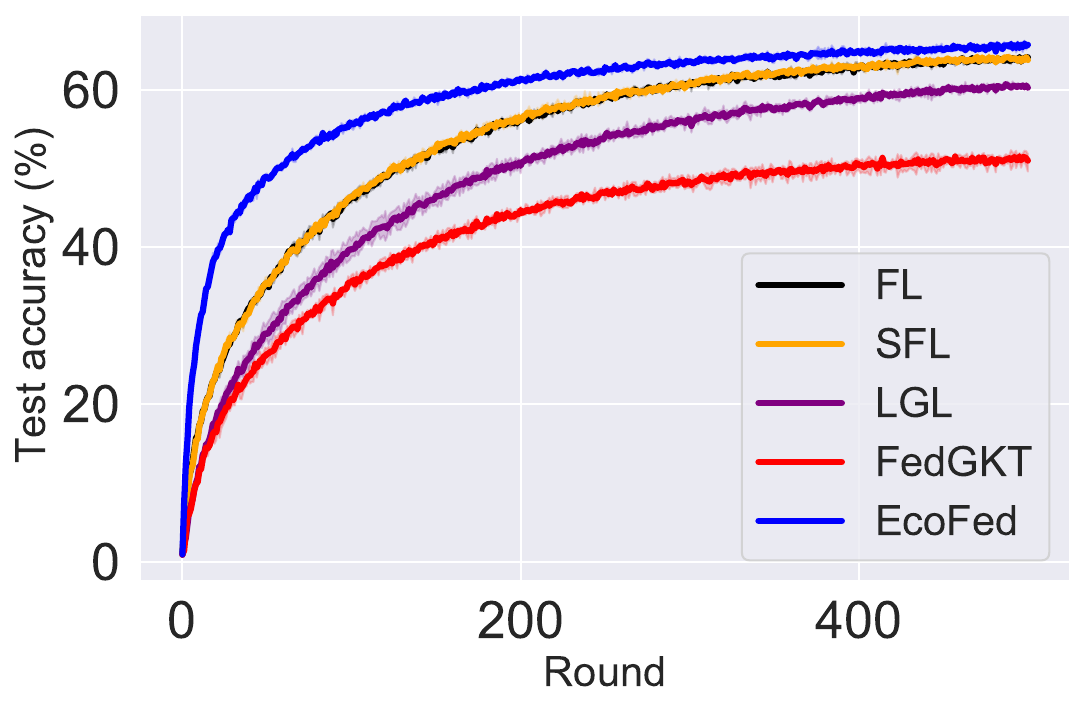}
         \caption{\centering ResNet9 on Non-I.I.D. \space CIFAR-100}
         \label{}
     \end{subfigure}
        \caption{Test accuracy curves of \mytitle and the baselines using VGG11 and ResNet9 in I.I.D. and Non-I.I.D. settings for CIFAR-10 and CIFAR-100 datasets.}
        \label{figure:convergence_curves}
        \squeezeup
\end{figure*}

%% file: Tables/acc.tex
\begin{table*}
	\centering
	\caption{The highest test accuracy of \mytitle compared to the baselines for VGG11 and ResNet9 on two datasets on I.I.D. and Non-I.I.D. distribution. The results are an average of three independent runs with different random seeds.}
    \begin{tabular}{ P{2cm} P{1.5cm} P{1.5cm} P{1.5cm} P{1.5cm} P{1.5cm} P{1.5cm} P{1.5cm} P{1.5cm}}
    \Xhline{2\arrayrulewidth}
    \multirow{3}{*}{\textbf{Methods}}& \multicolumn{4}{c}{\textbf{CIFAR-10}}&    \multicolumn{4}{c}{\textbf{CIFAR-100}}\\
    \cline{2-9}
     & \multicolumn{2}{c}{\textbf{I.I.D.}}& \multicolumn{2}{c}{\textbf{Non-I.I.D.}}&
     \multicolumn{2}{c}{\textbf{I.I.D.}}& \multicolumn{2}{c}{\textbf{Non-I.I.D.}}\\
     \cline{2-9}
     
     &
     \textbf{VGG11}&
     \textbf{ResNet9}&
     \textbf{VGG11}&
     \textbf{ResNet9}&
     \textbf{VGG11}&
     \textbf{ResNet9}&
     \textbf{VGG11}&
     \textbf{ResNet9}\\
     \hline
    FL&
    \cellcolor{lightgray!50} 88.29\%&
    \cellcolor{lightgray!50} 88.95\%& 
    \cellcolor{gray!50} 82.01\%&
    \cellcolor{gray!50} 84.52\%&
    \cellcolor{lightgray!50} 58.48\%&
    \cellcolor{lightgray!50} 65.87\%&
    \cellcolor{gray!50} 56.79\%&
    \cellcolor{gray!50} 64.29\%
    \\
    \hline
    SFL& 
    \cellcolor{lightgray!50} 88.31\%& 
    \cellcolor{lightgray!50} 88.81\%& 
    \cellcolor{gray!50} 81.38\%&  
    \cellcolor{gray!50} 84.56\%&   
    \cellcolor{lightgray!50} 58.5\%&
    \cellcolor{lightgray!50} 66.32\%&   
    \cellcolor{gray!50} 56.47\%&
    \cellcolor{gray!50} 64.42\%\\
   \hline
    LGL& 
    \cellcolor{lightgray!50} 85.76\%& 
    \cellcolor{lightgray!50} 85.65\%& 
    \cellcolor{gray!50} 79.66\%&   
    \cellcolor{gray!50} 82.18\%&
    \cellcolor{lightgray!50} 57.22\%&
    \cellcolor{lightgray!50} 63.01\%&   
    \cellcolor{gray!50} 54.88\%&
    \cellcolor{gray!50} 61\%\\
    \hline
    FedGKT& 
    \cellcolor{lightgray!50} 81.09\%& 
    \cellcolor{lightgray!50} 79.82\%& 
    \cellcolor{gray!50} 66.39\%&    
    \cellcolor{gray!50} 68.71\%&
    \cellcolor{lightgray!50} 52.03\%&
    \cellcolor{lightgray!50} 55.92\%&   
    \cellcolor{gray!50} 47.74\%&
    \cellcolor{gray!50} 51.81\%\\
    \hline
    \textbf{EcoFed}&  
    \cellcolor{lightgray!50} \textbf{88.87\%}& 
    \cellcolor{lightgray!50} \textbf{88.81\%}&  
    \cellcolor{gray!50} \textbf{84.47\%}&
    \cellcolor{gray!50} \textbf{85.82\%}&  
    \cellcolor{lightgray!50} \textbf{62.81\%}& 
    \cellcolor{lightgray!50} \textbf{67.61\%}&  
    \cellcolor{gray!50} \textbf{61.1\%}&  
    \cellcolor{gray!50} \textbf{66.08\%}\\

    \Xhline{2\arrayrulewidth}
    \end{tabular}
    \label{table:acc}
    \squeezeup
\end{table*}

%% file: Tables/fedgkt.tex
\begin{table}
	\centering
	\caption{The highest test accuracy of FedGKT (bi-direction) and FedGKT (uni-direction) on CIFAR-10. The results are an average of three independent runs with different random seeds.}
    \begin{tabular}{ P{2cm} P{1.1cm} P{1.1cm} P{1.1cm} P{1.1cm}}
    \Xhline{2\arrayrulewidth}
    \multirow{2}{*}{\textbf{FedGKT}}&   \multicolumn{2}{c}{\textbf{I.I.D.}}&  \multicolumn{2}{c}{\textbf{Non-I.I.D.}}\\
    \cline{2-5}
     &  \textbf{VGG11}&\textbf{ResNet9}&   \textbf{VGG11}&\textbf{ResNet9}\\
    \hline
    Bidirection&
    81.09\%&
    79.82\%&
    66.39\%&
    68.71\%\\

    Unidirection&
    83.18\%&
    83.42\%&
    75.89\%&
    77.68\%\\
    
    \Xhline{2\arrayrulewidth}
    \end{tabular}
    \label{table:fedgkt}
    \squeezeup
\end{table}

%% file: Tables/pretrained_init.tex
\begin{table}
	\centering
	\caption{The highest test accuracy of \mytitle compared to the baselines on CIFAR-10 with pre-trained initialization on the device-side model. The results are an average of three independent runs with different random seeds.}
    \begin{tabular}{ P{2cm} P{1.1cm} P{1.1cm} P{1.1cm} P{1.1cm}}
    \Xhline{2\arrayrulewidth}
    \multirow{2}{*}{\textbf{Methods}}&   \multicolumn{2}{c}{\textbf{I.I.D.}}&  \multicolumn{2}{c}{\textbf{Non-I.I.D.}}\\
    \cline{2-5}
     &  \textbf{VGG11}&\textbf{ResNet9}&   \textbf{VGG11}&\textbf{ResNet9}\\
    \hline
    FL&
    89.48\%&
    89.55\%&
    84.81\%&
    86.06\%\\

    SFL&
    89.44\%&
    89.57\%&
    85.02\%&
    86.19\%\\

    LGL&
    88.06\%&
    88.29\%&
    83.44\%&
    85.04\%\\

    FedGKT&
    83.59\%&
    82.35\%&
    72.73\%&
    74.36\%\\

    \hline
    \textbf{\mytitle}&  
    \textbf{88.87\%}&  
    \textbf{88.81\%}&  
    \textbf{84.47\%}&  
    \textbf{85.82\%}\\
    
    \Xhline{2\arrayrulewidth}
    \end{tabular}
    \label{table:pretrained_init}
    \squeezeup
\end{table}

%% file: Tables/lgl_frozon.tex
\begin{table}
	\centering
	\caption{The highest test accuracy of device-side model of trainable $\bm{w}_c$ and frozen $\bm{w}_c^*$ in LGL (on CIFAR-10) and the highest test accuracy of respective server-side models under $\bm{w}_c$ and $\bm{w}_c^*$. The results are an average of three independent runs with different random seeds.}
    \begin{tabular}{ P{0.4cm} P{1.8cm} P{1cm} P{1cm} P{1cm} P{1cm}}
    \Xhline{2\arrayrulewidth}
    \multicolumn{2}{c}{\multirow{2}{*}{\textbf{LGL}}}&   \multicolumn{2}{c}{\textbf{I.I.D.}}&  \multicolumn{2}{c}{\textbf{Non-I.I.D.}}\\
    \cline{3-6}
     & &  \textbf{VGG11}&\textbf{ResNet9}&   \textbf{VGG11}&\textbf{ResNet9}\\
    \hline

    \multirow{2}{*}{\textbf{Device}}&
    Trainable $\bm{w}_c$ &
    78.61\%&
    78.68\%&
    76.78\%&
    76.77\%\\
    
    &
    Frozen $\bm{w}_c^*$ &
    70.6\%&
    70.72\%&
    69.46\%&
    67.89\%\\

    \hline
    \multirow{2}{*}{\textbf{Server}}&
    $\bm{w}_s$ under $\bm{w}_c$ &
    87.82\%&
    88.04\%&
    83.05\%&
    84.9\%\\
    
    &
    $\bm{w}_s$ under $\bm{w}_c^*$ &
    88.97\%&
    89.34\%&
    84.4\%&
    85.77\%\\
    
    \Xhline{2\arrayrulewidth}
    \end{tabular}
    \label{table:lgl_frozon}
    \squeezeup
\end{table}

%% file: Results/dynamic_rho.tex
\begin{figure}
        \begin{subfigure}[b]{0.24\textwidth}
         \centering
         \includegraphics[width=\textwidth]{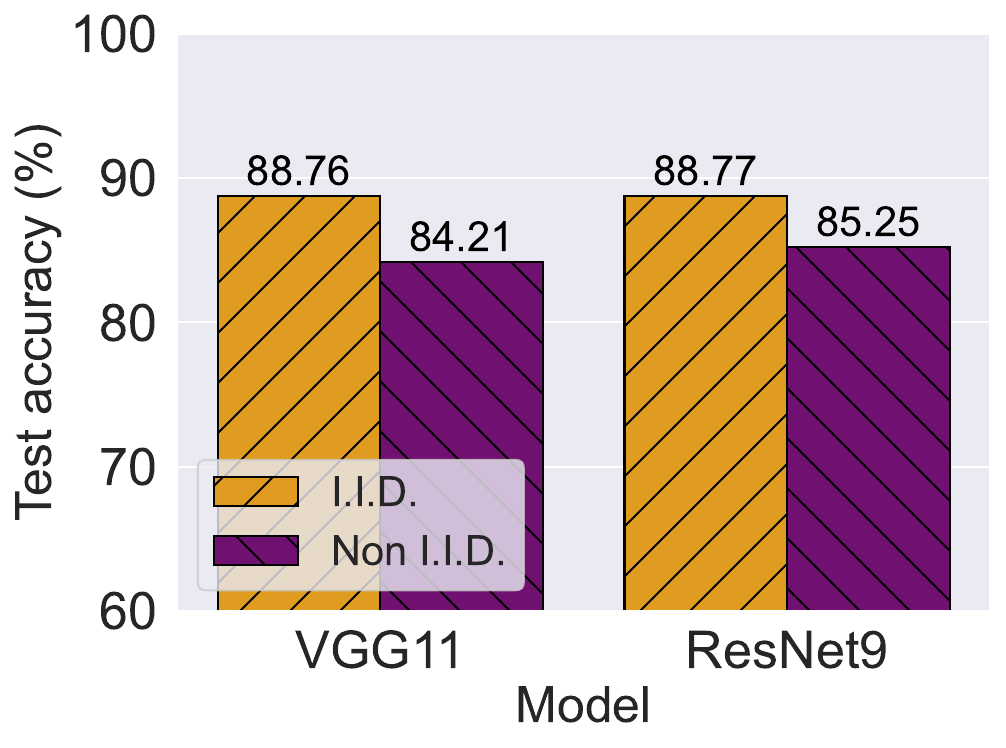}
         \caption{Accuracy}
         \label{figure:dynamic_rho_acc}
     \end{subfigure}
     \begin{subfigure}[b]{0.24\textwidth}
         \centering
         \includegraphics[width=\textwidth]{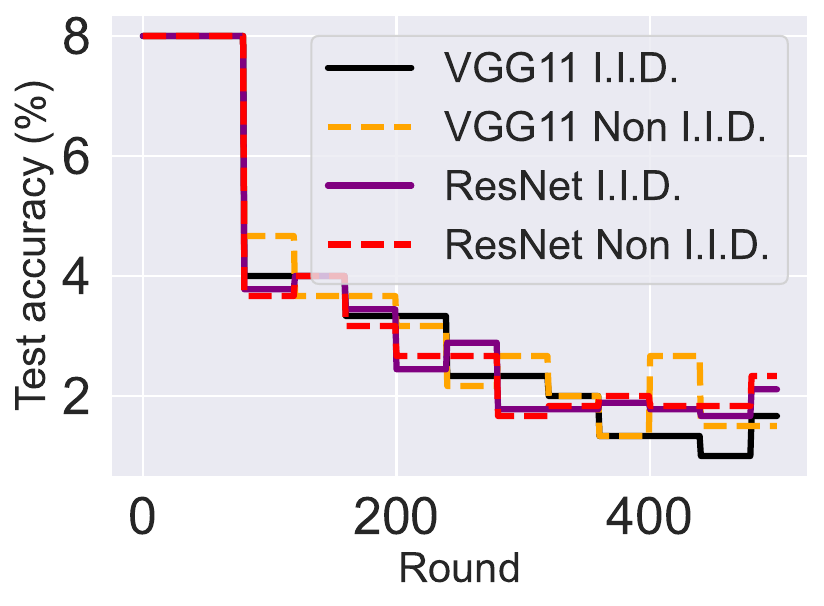}
         \caption{Period}
         \label{figure:dynamic_rho_period}
     \end{subfigure}
		\caption{\di{Dynamic $\rho$ values based on Table~\ref{table:dynamic_rho} in \mytitle. The results are an average of three independent runs with different random seeds.}}
		\label{fig:dynamic_rho}
\end{figure}

%% file: Results/period_quantization.tex
\begin{figure}[t]
        \begin{subfigure}[b]{0.24\textwidth}
         \centering
         \includegraphics[width=\textwidth]{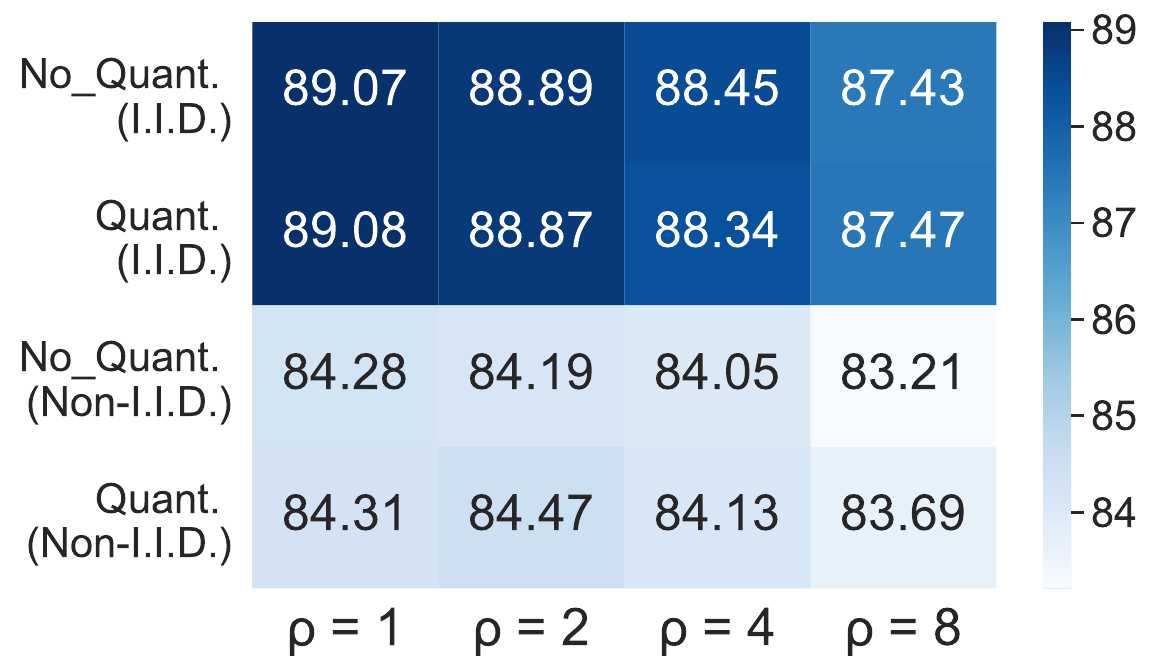}
         \caption{For VGG11}
         \label{figure:rho_quant_vgg11}
     \end{subfigure}
     \begin{subfigure}[b]{0.24\textwidth}
         \centering
         \includegraphics[width=\textwidth]{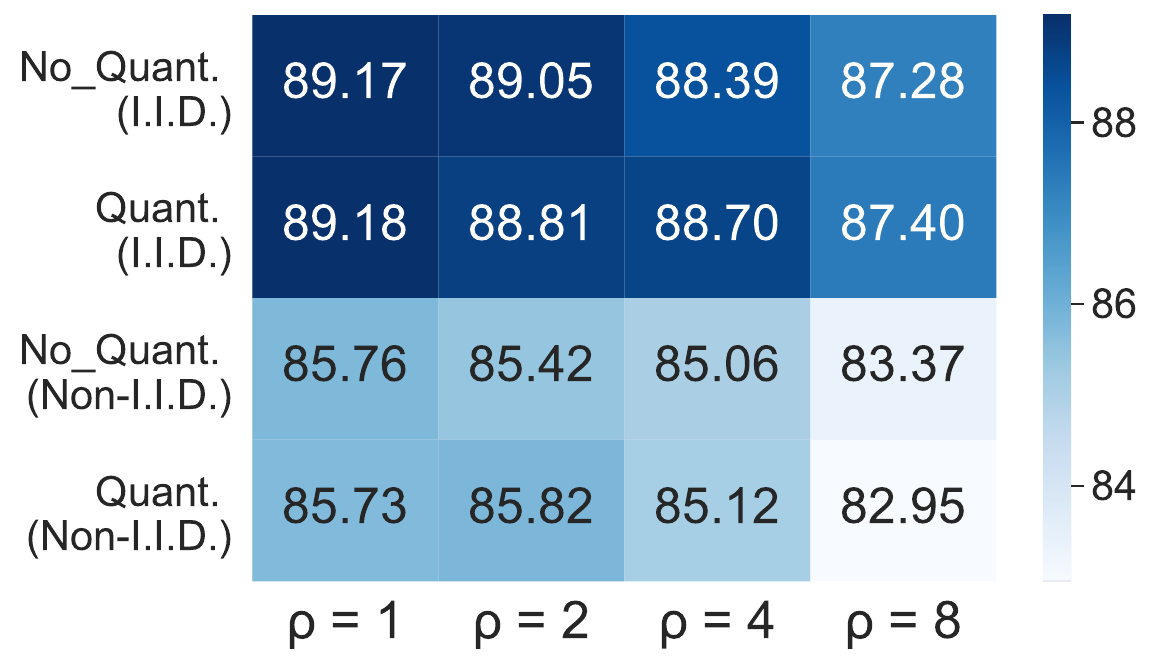}
         \caption{For ResNet9}
         \label{figure:rho_quant_resnet9}
     \end{subfigure}
		\caption{Accuracy for different $\rho$ values with or without quantization in \mytitle. The results are an average of three independent runs with different random seeds.}
		\label{fig:period_quantization}
\end{figure}

%% file: Tables/pretraining_datasets.tex
\begin{table}[t]
	\centering
	\caption{\di{The highest test accuracy of \mytitle on CIFAR-10 with different pre-training datasets. We use \textbf{L} and \textbf{S} to denote large-scale or small-scale datasets, and \textbf{Nat} and \textbf{Syn} to denote whether the data is natural or synthetic. The results are an average of three independent runs with different random seeds.}}
    \di{
    \begin{tabular}{ P{3cm} P{0.9cm} P{0.9cm} P{0.9cm} P{0.9cm}}
    \Xhline{2\arrayrulewidth}
    \multirow{2}{*}{\textbf{Pre-training datasets}}&   \multicolumn{2}{c}{\textbf{I.I.D.}}&  \multicolumn{2}{c}{\textbf{Non-I.I.D.}}\\
    \cline{2-5}
     &  \textbf{VGG11}&\textbf{ResNet9}&   \textbf{VGG11}&\textbf{ResNet9}\\
    \hline
    ImageNet (L, Nat)&
    88.87\%&
    88.81\%&
    84.47\%&
    85.82\%\\

    Tiny-ImageNet (S, Nat)&
    89.32\%&
    88.99\%&
    85.66\%&
    85.86\%\\

    CIFAR-5m (S, Syn)&
    87.92\%&
    87.94\%&
    83.61\%&
    83.6\%\\

    SIP-17 (S, Syn)&
    86.81\%&
    86.98\%&
    81.45\%&
    83.49\%\\
    
    \Xhline{2\arrayrulewidth}
    \end{tabular}
    }
    \label{table:pretraining_datasets}
    \squeezeup
\end{table}

%% file: Tables/comm.tex
\begin{table}
	\centering
	\caption{Communication cost for one training round.}
    \begin{tabular}{ P{3cm} P{1.5cm} P{1.5cm} }
    \Xhline{2\arrayrulewidth}
    \multirow{2}{*}{\textbf{Methods}}& 
    \multicolumn{2}{c}{\textbf{Communication cost}}\\
    \cline{2-3}

     &  VGG11&ResNet9\\
     \hline
     FL& 5.13 GB& 1.4 GB\\

     SFL& 0.62 GB& 0.62 GB\\

     LGL& 0.33 GB& 0.33 GB\\

     FedGKT& 0.33 GB& 0.33 GB\\
     \hline
     \textbf{\mytitle w/o buffer}&  \textbf{\di{0.077 GB}}&  \textbf{\di{0.077 GB}}\\

     \textbf{\mytitle w buffer}&  \textbf{0 GB}&  \textbf{0 GB}\\

     \Xhline{2\arrayrulewidth}
    \end{tabular}
	\label{table:comm}
    \squeezeup
\end{table}

%% file: Results/latency.tex
\begin{figure}
     \centering
     \begin{subfigure}[b]{0.24\textwidth}
         \centering
         \includegraphics[width=\textwidth]{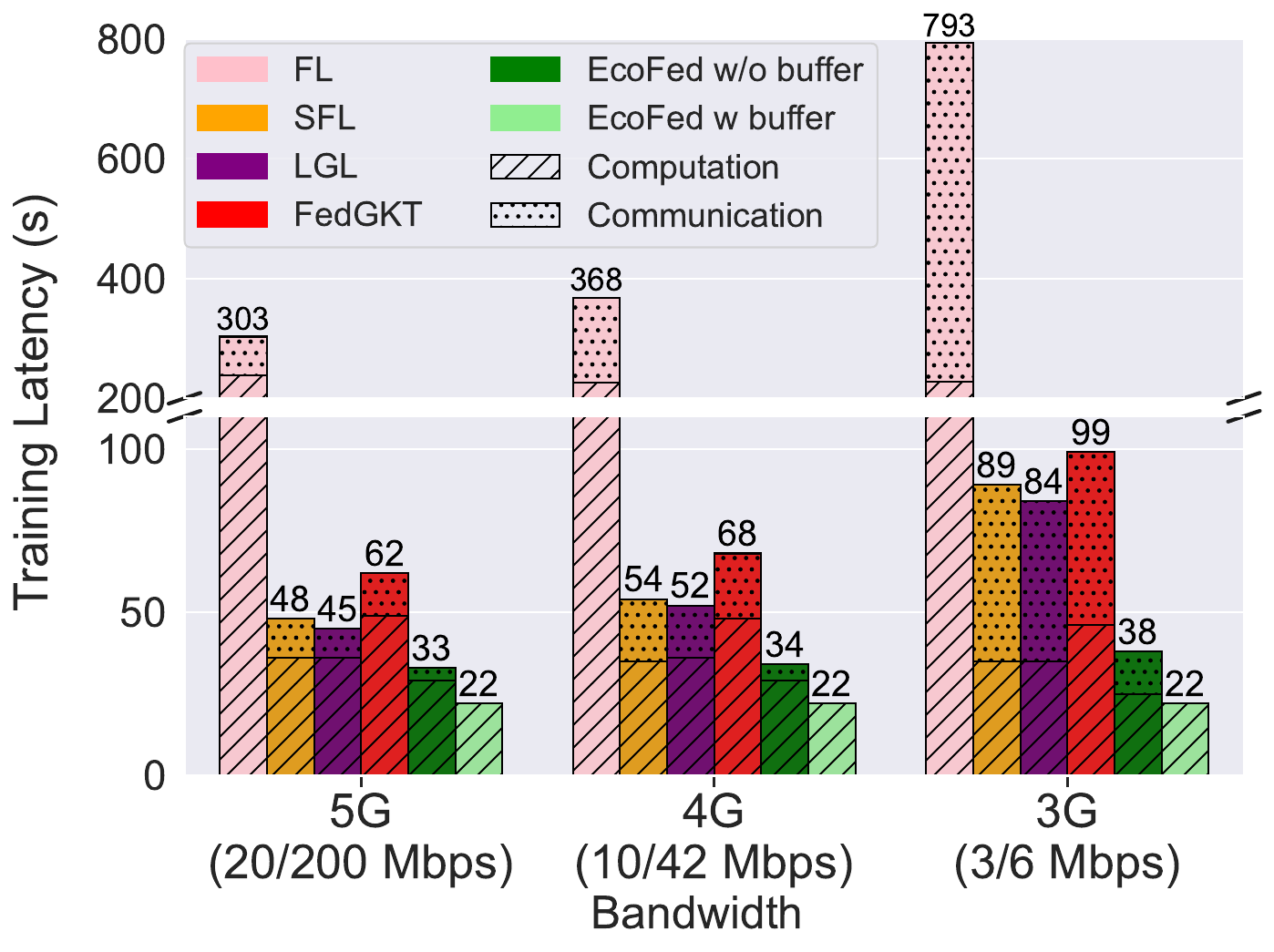}
         \caption{For VGG11}
         \label{figure:latency_vgg11}
     \end{subfigure}
     \begin{subfigure}[b]{0.24\textwidth}
         \centering
         \includegraphics[width=\textwidth]{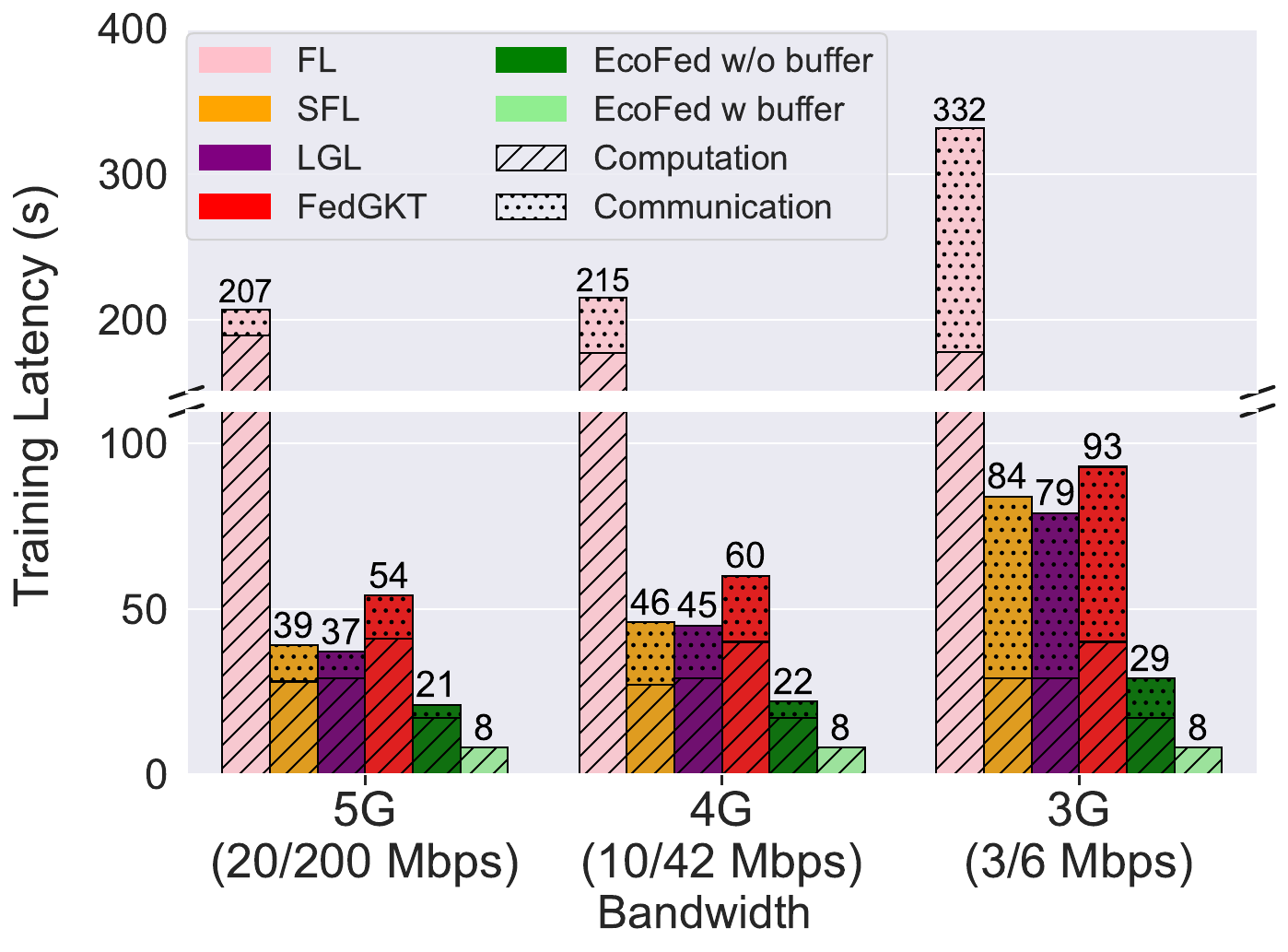}
         \caption{For ResNet9}
         \label{figure:latency_resnet9}
     \end{subfigure}
        \caption{\di{Latency of one training round for VGG11 and ResNet9 under different network conditions for PP2. The results are an average of three independent runs.}}
        \label{figure:latency}
        \squeezeup
\end{figure}

%% file: Results/latency_pps.tex
\begin{figure}
     \centering
     \begin{subfigure}[b]{0.24\textwidth}
         \centering
         \includegraphics[width=\textwidth]{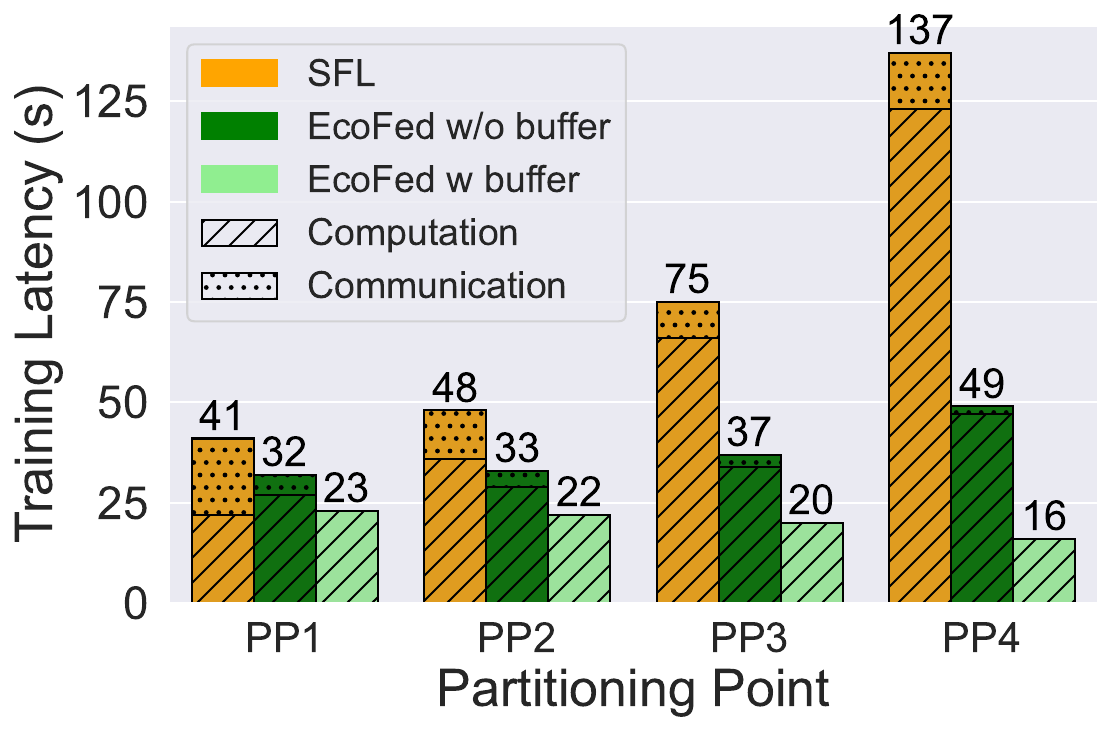}
         \caption{For VGG11}
         \label{figure:latency_pps_vgg11}
     \end{subfigure}
     \begin{subfigure}[b]{0.24\textwidth}
         \centering
         \includegraphics[width=\textwidth]{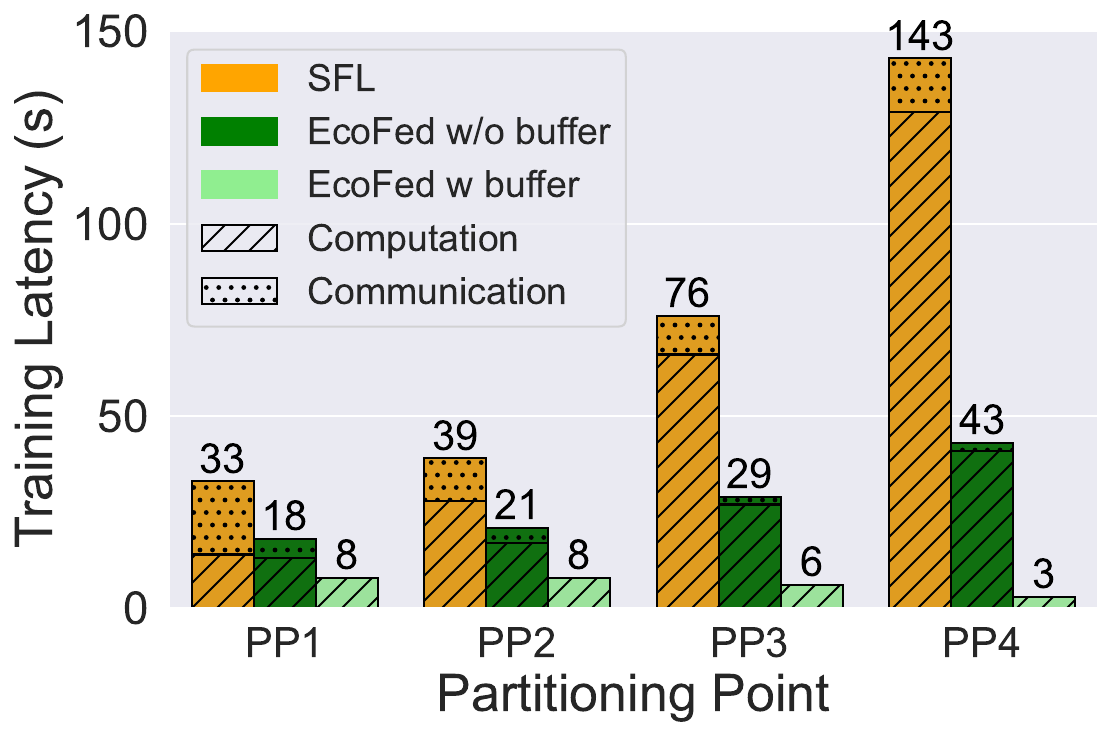}
         \caption{For ResNet9}
         \label{figure:latency_pps_resnet9}
     \end{subfigure}
        \caption{\di{Latency of one training round for VGG11 and ResNet9 under different partitioning points in 5G conditions. The results are an average of three independent runs.}}
        \label{figure:latency_pps}
        \squeezeup
\end{figure}

%% file: Results/comm_vs_acc.tex
\begin{figure*}
     \centering
     \begin{subfigure}[b]{0.24\textwidth}
         \centering
         \includegraphics[width=\textwidth]{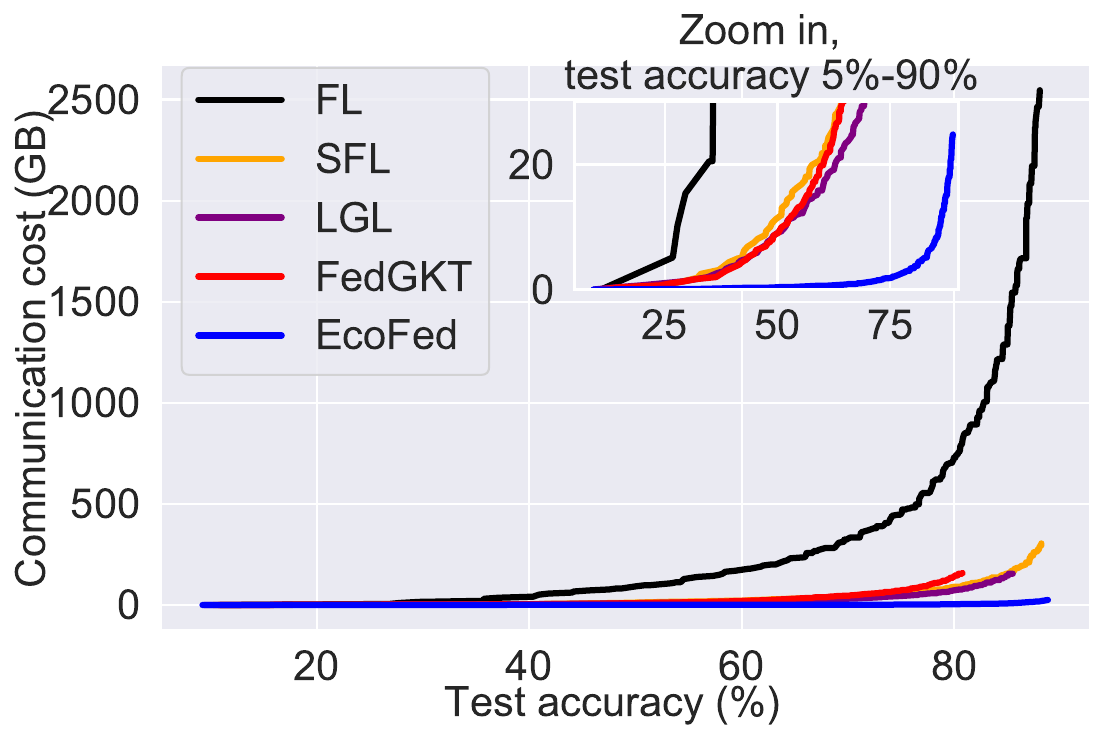}
         \caption{VGG11 on I.I.D. CIFAR-10}
         \label{}
     \end{subfigure}
     \hfill
     \begin{subfigure}[b]{0.24\textwidth}
         \centering
         \includegraphics[width=\textwidth]{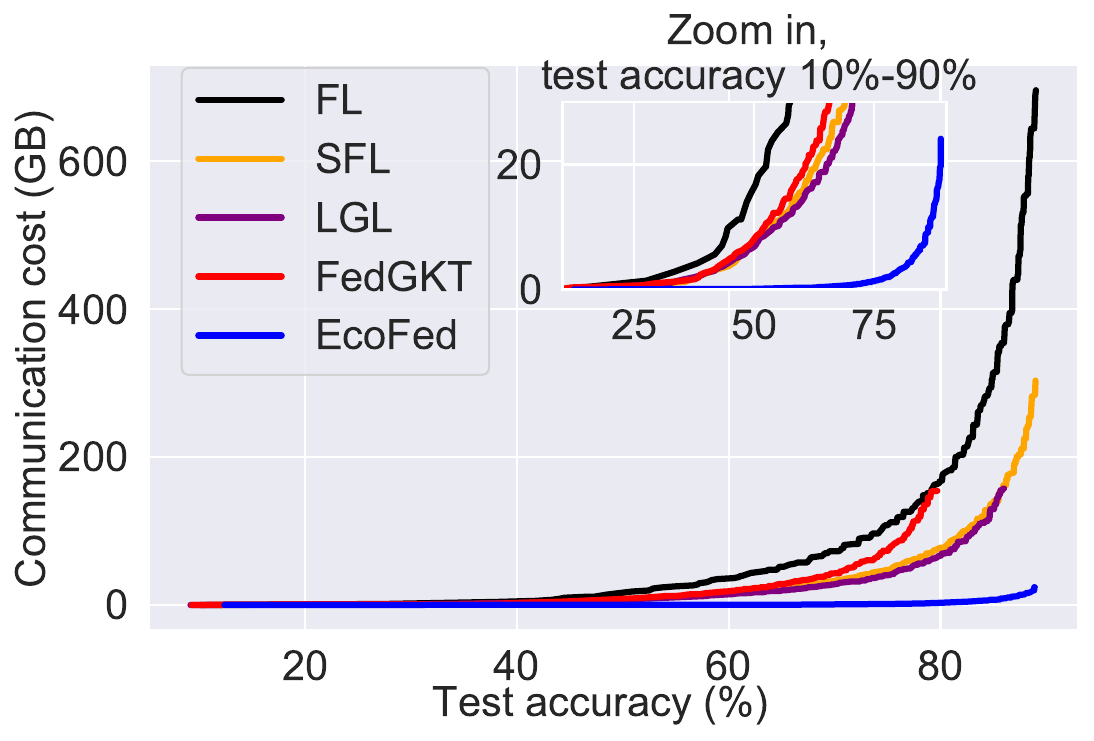}
         \caption{ResNet9 on I.I.D. CIFAR-10}
         \label{}
     \end{subfigure}
     \hfill
     \begin{subfigure}[b]{0.24\textwidth}
         \centering
         \includegraphics[width=\textwidth]{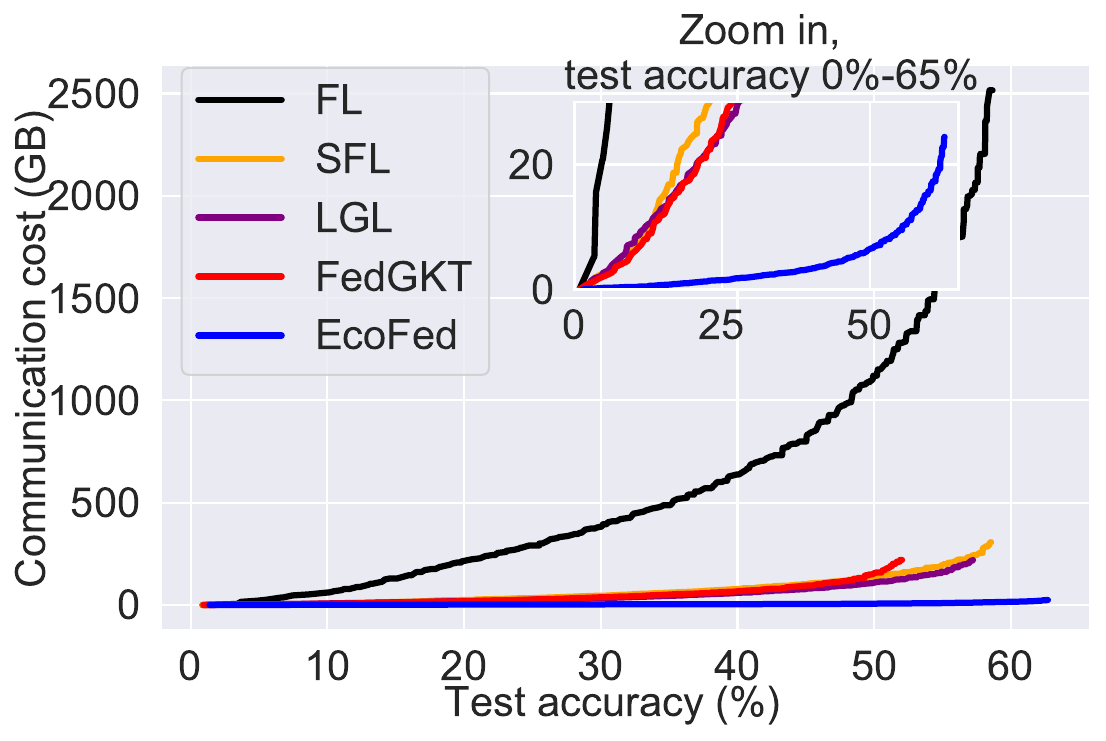}
         \caption{VGG11 on I.I.D. CIFAR-100}
         \label{}
     \end{subfigure}
     \hfill
     \begin{subfigure}[b]{0.244\textwidth}
         \centering
         \includegraphics[width=\textwidth]
         {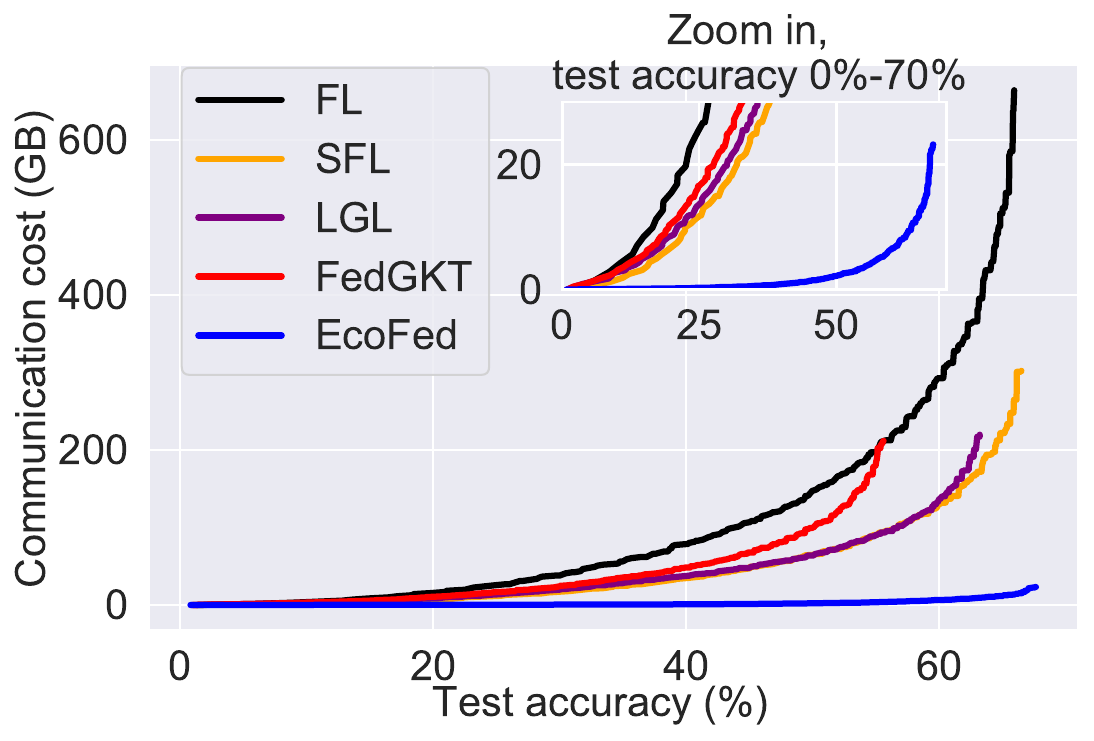}
         \caption{ResNet9 on I.I.D. CIFAR-100}
         \label{}
     \end{subfigure}
     \hfill
     \begin{subfigure}[b]{0.24\textwidth}
         \centering
         \includegraphics[width=\textwidth]{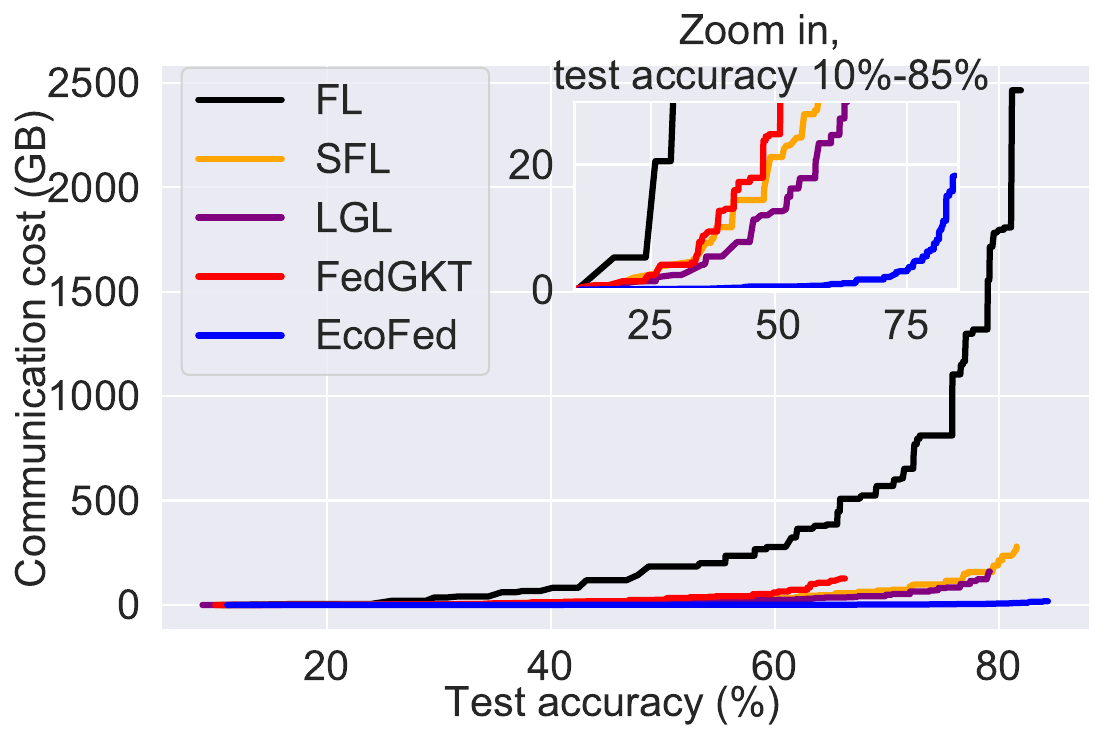}
         \caption{\centering VGG11 on Non-I.I.D. CIFAR-10}
         \label{}
     \end{subfigure}
     \hfill
     \begin{subfigure}[b]{0.24\textwidth}
         \centering
         \includegraphics[width=\textwidth]{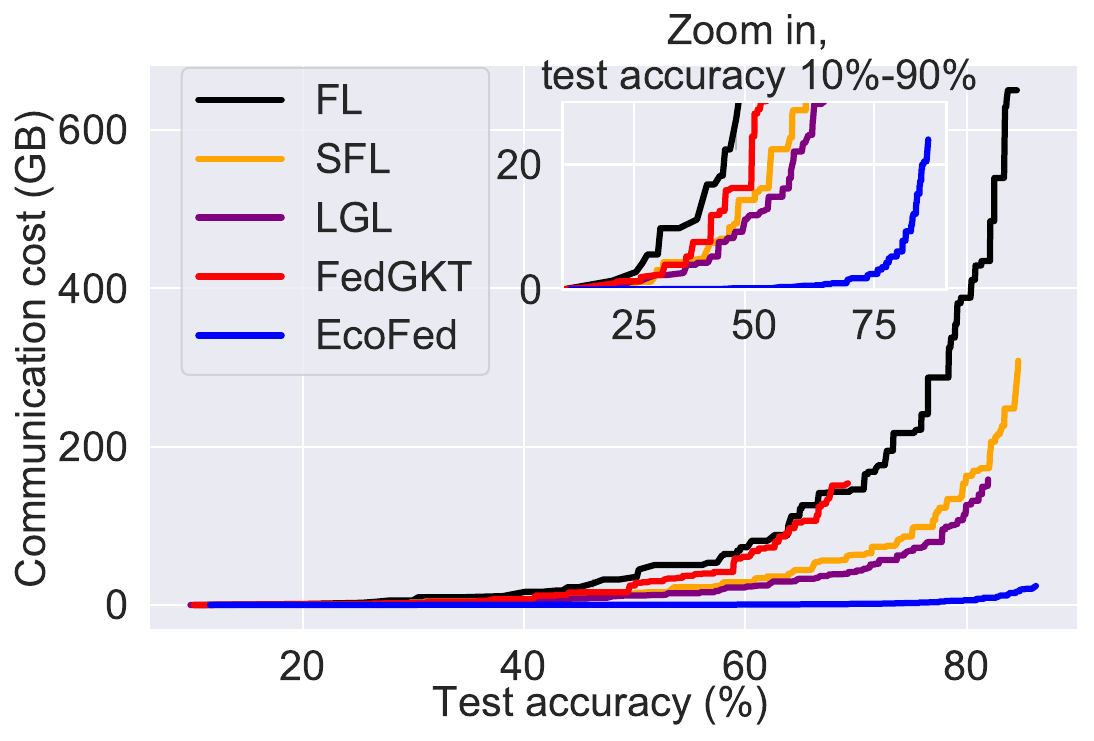}
         \caption{\centering ResNet9 on Non-I.I.D. CIFAR-10}
         \label{}
     \end{subfigure}
     \hfill
     \begin{subfigure}[b]{0.24\textwidth}
         \centering
         \includegraphics[width=\textwidth]{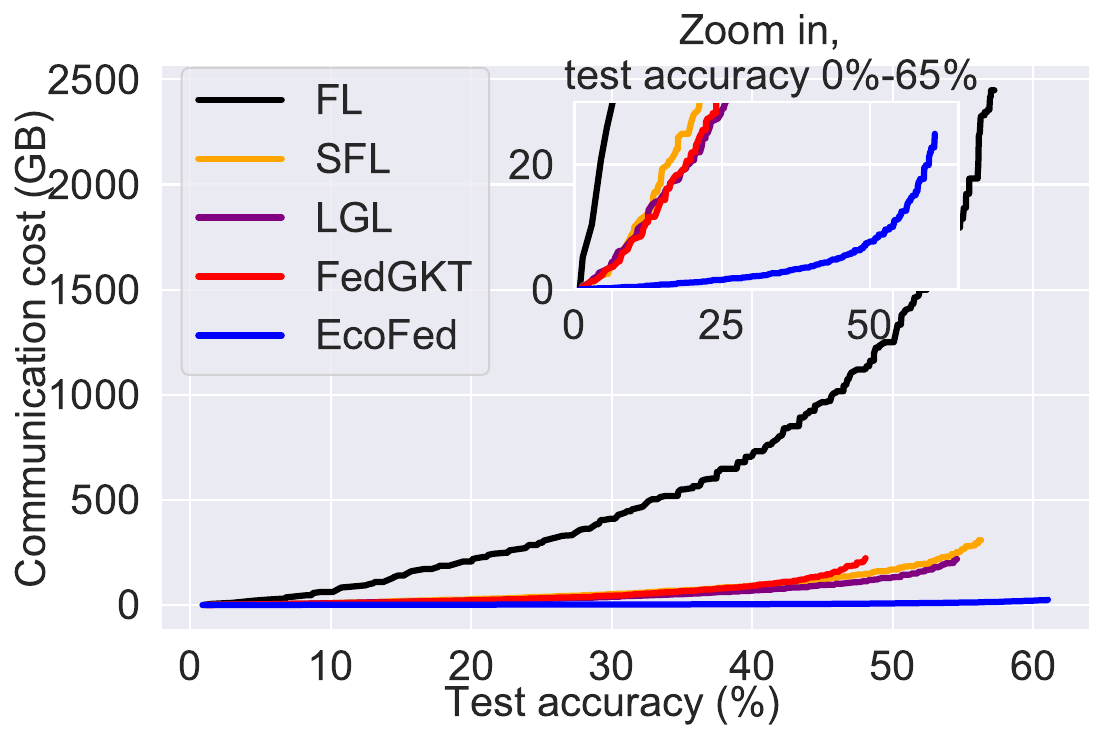}
         \caption{\centering VGG11 on Non-I.I.D. \space CIFAR-100}
         \label{}
     \end{subfigure}
     \hfill
     \begin{subfigure}[b]{0.24\textwidth}
         \centering
         \includegraphics[width=\textwidth]{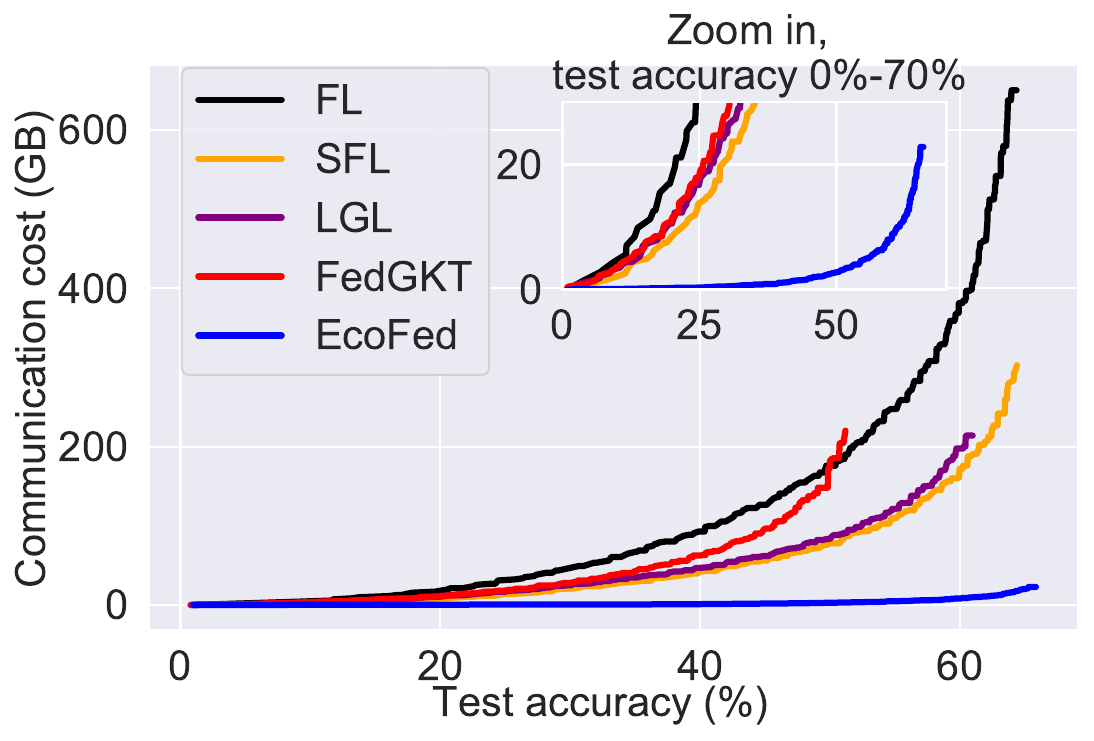}
         \caption{\centering ResNet9 on Non-I.I.D. \space CIFAR-100}
         \label{}
     \end{subfigure}
        \caption{Communication cost versus test accuracy of VGG11 and ResNet9 for the I.I.D and Non-I.I.D. settings on CIFAR-10 and CIFAR-100 datasets.}
        \label{figure:commvsacc}
\end{figure*}

%% file: Sections/related_work.tex
\di{
In this section, we consider techniques for reducing communication in FL and the literature on DPFL methods. 
We also discuss pre-training in FL and layer-wise learning.
}

\begin{table*}
\begin{center}
\caption{\di{Comparing FL, vanilla DPFL, local loss-based  DPFL and \mytitle.}}
\di{
\begin{tabular}{ P{7.5cm} P{1cm} P{3cm} P{3cm} P{1.5cm}}
\Xhline{2\arrayrulewidth}
& FL & Vanilla DPFL (e.g.,~\cite{thapa2022splitfed,he2023accelerated,deng2022low,wu2022fedadapt}) & Local loss-based DPFL (e.g.,~\cite{hanaccelerating,he2020group}) & \mytitle \\
\hline
Offers device-side acceleration  & \xmark  & \cmark  & \cmark & \cmark\\
\hline
Reduces communication cost for DNN partitioning & n/a & \xmark  & \cmark & \cmark\\
\hline
Optimizes communication arising from activation transfers & n/a & \xmark  & \xmark & \cmark\\
\hline
Has low accuracy degradation & n/a & n/a  & \xmark & \cmark\\
\Xhline{2\arrayrulewidth}
\end{tabular}
}
\label{table:works}
\end{center}
\end{table*}

\textit{\di{Reducing} Communication in FL}:
Communication reduction techniques in FL can be grouped into two categories, depending on whether they (i) reduce the frequency of communication or (ii) compress the size of the transferred data. Under the first category, a key technique is to increase the interval between model aggregation, thus reducing the communication frequency from the device to the server and vice-versa~\cite{wang2019adaptive,wu2023hiflash}. Under the second category, compression approaches, such as quantization and sparsification, are employed to minimize the size of models (updated weights) in each round of communication~\cite{reisizadeh2020fedpaq,han2020adaptive}. \di{By incorporating distillation, the communication overhead of transferring model parameters in traditional FL can be eliminated~\cite{itahara2021distillation}.}
However, it is the above focus on the communication of the updated models at the end of each FL round, rather than the communication costs introduced due to DNN partitioning during training, that our \di{article} considers.

\textit{DPFL}: 
\di{Existing research on DPFL can be categorized as vanilla DPFL and local loss-based DPFL based on how they optimize the communication.} SFL~\cite{thapa2022splitfed} is the first DPFL work that combines FL and split learning~\cite{vepakomma2018split} such that the device-side models are independently trained by receiving gradients from the server. The server-side models are trained by collecting activations from devices in parallel. \di{In addition, dynamic partitioning strategies for DPFL~\cite{he2023accelerated,deng2022low,wu2022fedadapt} and pipeline scheduling~\cite{zhang2022pipelearn} have been considered to optimize performance. However, these methods are considered vanilla DPFL as they do not consider the communication overhead introduced by DNN partitioning.}

Recently DPFL approaches have been optimized by computing local loss on the device-side to reduce the communication cost~\cite{hanaccelerating,he2020group}. In these approaches, the device-side model is trained with local error signals generated by an additional auxiliary network. This eliminates the need for transferring the gradient from the server and making use of it on the device. However, the training of device-side model with local error signals is sub-optimal, and thus detrimental to accuracy. In addition, the communication costs to transfer the activation are not considered. \di{
We highlight the differences between classic FL, vanilla DPFL, local loss-based DPFL and \mytitle in Table~\ref{table:works}.
}

\textit{Pre-training in FL}:
Pre-training a model is rarely investigated in the literature on FL. Instead, the model is usually trained from random weights. Recent work has demonstrated that pre-training can close the accuracy gap between FL and centralized learning, specifically in the non-IID setting~\cite{chen2022pre,nguyen2022begin}. \mytitle utilizes pre-training on the device-side and presents an approach to reduce the accuracy loss inherent to local loss-based DPFL methods; both of these are considered for the first time. 

\textit{Layer-wise Learning}:
Another line of related research is layer-wise learning in which each DNN layer is independently trained using auxiliary networks~\cite{nokland2019training,belilovsky2019greedy,belilovsky2020decoupled}. Since these approaches train using local loss in a resource-rich environment (with centralized servers that have an abundance of resources), the computation overheads of each layer and communication costs to transfer intermediate data are rarely considered. \mytitle can be considered as a special case of parallel block-wise learning. However, the device-side model is deployed in a resource-constrained environment on devices with limited computation and communication resources, which is challenging.

%% file: Sections/conclusion.tex
In this \di{article}, we present \mytitle, a communication-efficient DNN partitioning-based federated learning (DPFL) framework. DPFL partitions a DNN and offloads some of the layers of the DNN (or computation) from a resource constrained device to the server. \mytitle proposes pre-trained initialization to eliminate the transmission of the gradient from the server-side model to device-side model in DPFL for the first time and designs a novel replay buffer mechanism with quantization-based compression to further reduce the communication cost incurred by transferring activation. In other words, \mytitle proposes a unique way of carrying out the forward and backward passes for efficiently training in resource constrained environments. We comprehensively evaluate \mytitle and demonstrate that \mytitle can reduce the accuracy degradation caused by state-of-the-art local loss-based DPFL methods while significantly improving the communication efficiency and training speed compared to classical FL and state-of-the-art DPFL methods.

%% file: Sections/acknowledgement.tex
This work was sponsored by Rakuten Mobile, Japan. 

%% file: Sections/Authorsbiography.tex
\vskip -2.5\baselineskip plus -1fil
\begin{IEEEbiography}[{\includegraphics[width=1in,height=1.25in,clip,keepaspectratio]{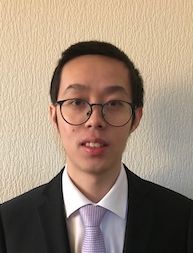}}]{Di Wu} is currently pursuing a PhD degree in computer science at University of St Andrews, UK. He received a B.S. degree in Information System and Information Management from Northeast Forestry University, China in 2015, and an M.S. degree in Data Science from University of Southampton, UK in 2018. His major interests are in the areas of federated learning, distributed machine learning, edge computing, model compression, and Internet-of-Things.
\end{IEEEbiography}
\vskip -2.5\baselineskip plus -1fil

\begin{IEEEbiography}[{\includegraphics[width=1in,height=1.25in,clip,keepaspectratio]{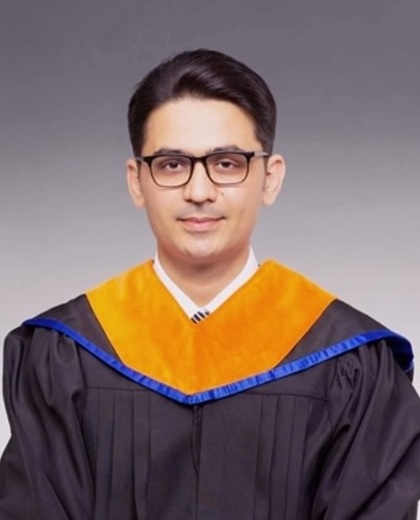}}]{Rehmat Ullah} earned a PhD degree in electronics and computer engineering from Hongik University, South Korea. He is currently an Assistant Professor at the Cardiff School of Technologies, Cardiff Metropolitan University, UK. Previously, he worked as a research fellow at University of St Andrews, UK and as an assistant professor at Gachon University, South Korea. His research interests are in edge computing, information centric networking and 5G evolution and beyond with a recent focus on federated learning for edge computing systems.  More information is available from \url{www.rehmatkhan.com}.
\end{IEEEbiography}
\vskip -2.5\baselineskip plus -1fil

\begin{IEEEbiography}[{\includegraphics[width=1in,height=1.25in,clip,keepaspectratio]{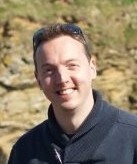}}]{Philip Rodgers}
received the PhD degree in Computer Science from the University of Strathclyde, UK in 2019. He is currently a Research Scientist at the Rakuten Mobile Innovation Studio, Japan. His research interests are in distributed algorithms for artificial intelligence. 
\end{IEEEbiography}
\vskip -2.5\baselineskip plus -1fil

\begin{IEEEbiography}[{\includegraphics[width=1in,height=1.25in,clip,keepaspectratio]{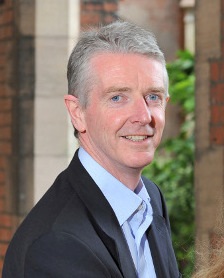}}]{Peter Kilpatrick} received the PhD degree in Computer Science in 1985, and the BSc degree in  Mathematics and Computer Science in 1981. He is currently a Reader in Computer Science at Queen's University Belfast. His interests include parallel programming models and cloud and edge computing.
\end{IEEEbiography}
\vskip -2.5\baselineskip plus -1fil

\begin{IEEEbiography}[{\includegraphics[width=1in,height=1.25in,clip,keepaspectratio]{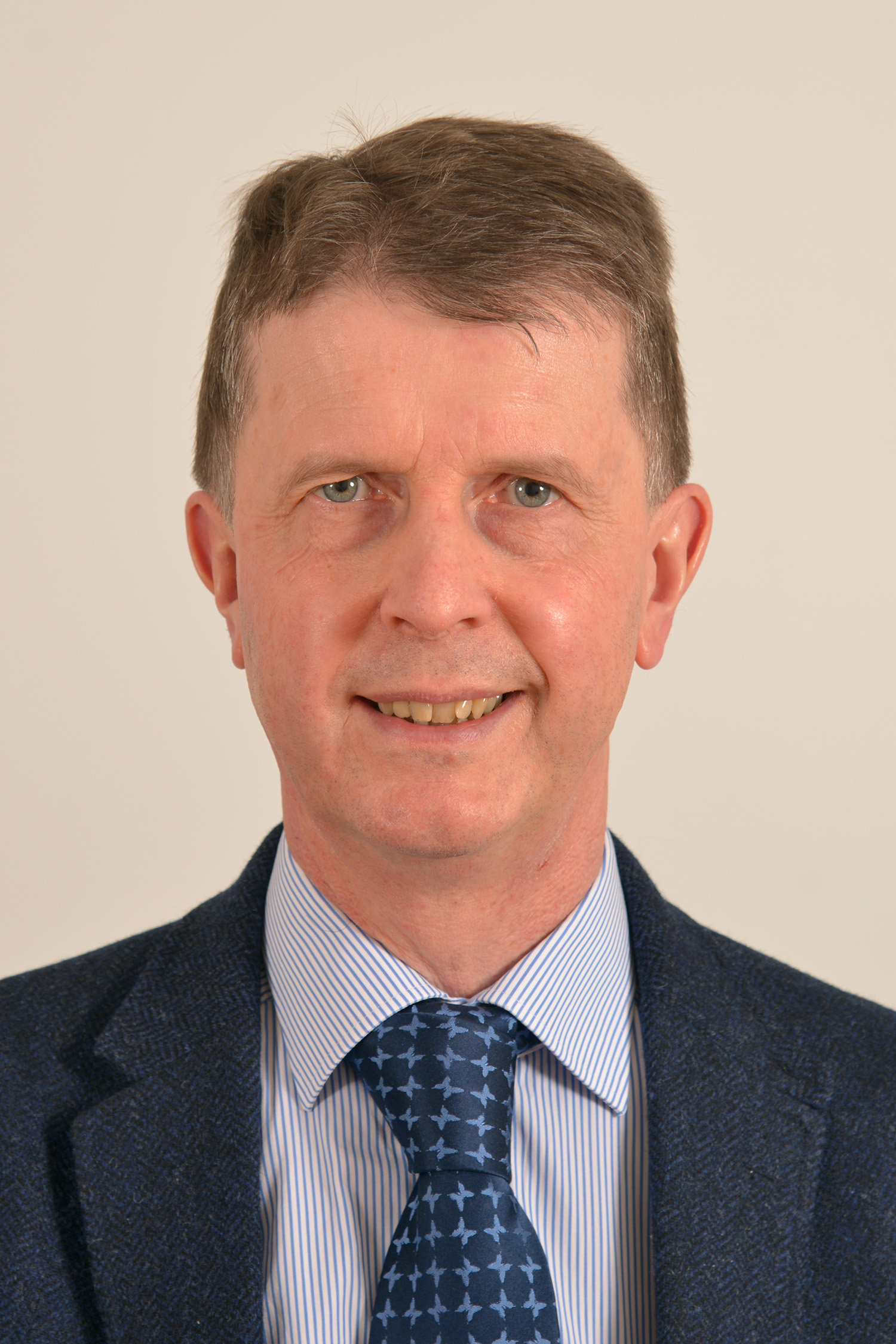}}]{Ivor Spence} received the PhD degree in computer science from Queen's University Belfast, UK, where he did research on code generation. He is currently a Reader in computer science at Queen's University Belfast where he leads the artificial intelligence (AI) research theme.
His research is primarily on heterogeneous computing systems for AI.
\end{IEEEbiography}
\vskip -2.5\baselineskip plus -1fil

\begin{IEEEbiography}[{\includegraphics[width=1in,height=1.25in,clip,keepaspectratio]{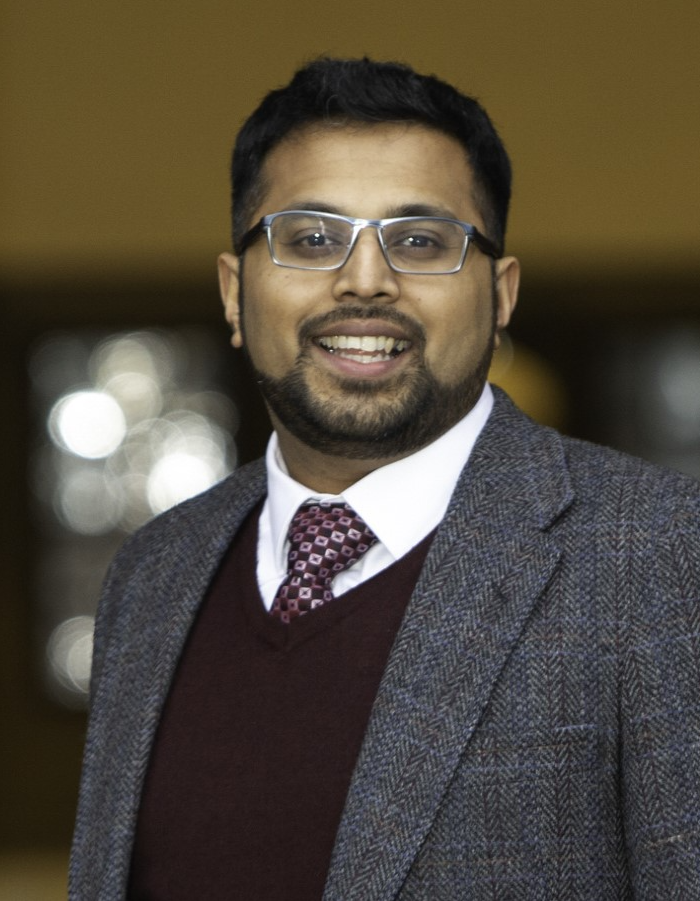}}]{Blesson Varghese} received the PhD degree in Computer Science from the University of Reading, UK on international scholarships. He is a Reader in Computer Science at the University of St Andrews, UK, and the Principal Investigator of the Edge Computing Hub. He is a previous Royal Society Short Industry Fellow. His interests include distributed systems that span the cloud-edge-device continuum and edge intelligence applications. More information is available from \url{www.blessonv.com}.
\end{IEEEbiography}

%% file: Sections/Appendix_Trans.tex
\newpage
\thispagestyle{empty}
\appendices
\section{Convergence of the Server-side Model}
\label{appendix:proof}
This Appendix presents the proof to demonstrate the convergence of the server-side model ($\bm{w}_{S}$).
\begin{align*}
\frac{1}{\Gamma_T} \sum_{t=0}^{T-1} \eta_t \mathbb{E} [ \lVert \nabla F_S(\bm{w}_S^{t}) \rVert^2 ] 
\leq 
\frac{4 ( F_S(\bm{w}_S^{0}) - F_S(\bm{w}_S^{*}) )}{3 \Gamma_T} \\
+ 
\frac{1}{\Gamma_T} \sum_{t=0}^{T-1} \left( \eta_t \left(\sqrt{G} + 1\right) \left( H_1+H_2 \right) + \frac{L}{2} \eta_t^2 G \right)
\tag{1}
\label{eq1}
\end{align*}

\noindent Based on Assumption 1 presented in the article, we have
\begin{align*}
F_S(\bm{w}_S^{t+1}) \leq F_S(\bm{w}_S^{t}) + \nabla F_S(\bm{w}_S^{t})^T(\bm{w}_S^{t+1}-\bm{w}_S^{t}) \\
+ \frac{L}{2} \lVert \bm{w}_S^{t+1}-\bm{w}_S^{t} \rVert^2
\tag{2}
\label{eq2}
\end{align*}

\noindent Also, note that we have
\begin{align*}
\bm{w}_S^{t+1} = \bm{w}_S^{t} - \eta_t \frac{1}{K} \sum_{k=1}^K {\nabla} F_{S,k}(\bm{w}_S^{t})
\tag{3}
\label{eq3}
\end{align*}

\noindent When substituting Equation~\ref{eq3} in Equation~\ref{eq2} we have
\begin{align*}
F_S(\bm{w}_S^{t+1}) 
\leq F_S(\bm{w}_S^{t}) \\
- \eta_t \nabla F_S(\bm{w}_S^{t})^T \left( \frac{1}{K} \sum_{k=1}^K {\nabla} F_{S,k}(\bm{w}_S^{t}) \right) \\
+ \frac{L}{2} \eta_t^2 \left\lVert \frac{1}{K} \sum_{k=1}^K {\nabla} F_{S,k}(\bm{w}_S^{t}) \right\rVert ^2
\tag{4}
\label{eq4}
\end{align*}

\noindent By taking the expectation from both sides of Equation~\ref{eq4} we have
\begin{align*}
\mathbb{E}[ F_S(\bm{w}_S^{t+1}) ] 
\leq \mathbb{E}[ F_S(\bm{w}_S^{t}) ] \\
- \eta_t \underbrace{ \mathbb{E} \left[ \nabla F_S(\bm{w}_S^{t})^T \left( \frac{1}{K} \sum_{k=1}^K {\nabla} F_{S,k}(\bm{w}_S^{t}) \right) \right]
}_{B_1} \\
+ \frac{L}{2} \eta_t^2 \underbrace{ \mathbb{E}\left[ \left\lVert \frac{1}{K} \sum_{k=1}^K 
{\nabla} F_{S,k}(\bm{w}_S^{t}) \right\rVert ^2 \right]
}_{B_2}
\tag{5}
\label{eq5}
\end{align*}

\noindent Next, we will find the lower bound of $B_1$ and upper bound of $B_2$. We define $X$ as
\begin{align*}
X = \frac{1}{K} \sum_{k=1}^K ({\nabla} F_{S,k}(\bm{w}_S^{t};\bm{a}_k^t) - \nabla F_{S,k}(\bm{w}_S^{t};\hat{\bm{a}}_k^t))
\tag{6}
\label{eq6}
\end{align*}

\noindent which is the difference in average gradient using $\bm{a}_k^t$ and $\hat{\bm{a}}_k^t$. For simplicity, we use the following abbreviation.
\begin{align*}
X = \frac{1}{K} \sum_{k=1}^K (\nabla F_{S,k}(\bm{w}_S^{t}) - \hat{\nabla} F_{S,k}(\bm{w}_S^{t}))
\tag{7}
\label{eq7}
\end{align*}

\noindent Substituting $X$ for $B_1$ we have
\begin{align*}
&B_1 = \mathbb{E} \left[ \nabla F_S(\bm{w}_S^{t})^T \left( \frac{1}{K} \sum_{k=1}^K {\nabla} F_{S,k}(\bm{w}_S^{t}) \right) \right] \tag{8} \\
&= \mathbb{E} \left[ \nabla F_S(\bm{w}_S^{t})^T \left( X + \frac{1}{K} \sum_{k=1}^K \hat{\nabla} F_{S,k}(\bm{w}_S^{t}) \right) \right] \tag{9} \\
&\geq \mathbb{E} \left[ \nabla F_S(\bm{w}_S^{t})^T \left( \frac{1}{K} \sum_{k=1}^K \hat{\nabla} F_{S,k}(\bm{w}_S^{t}) \right) \right] \\
&- \lVert  \mathbb{E} [ \nabla F_S(\bm{w}_S^{t})^T X ] \rVert \tag{10} \\
&= \underbrace{ \mathbb{E} \left[ \nabla F_S(\bm{w}_S^{t})^T \left( \frac{1}{K} \sum_{k=1}^K \nabla F_{S,k}(\bm{w}_S^{t}) \right) \right] }_{C_1} \\
&+ \underbrace{ \mathbb{E} \left[ \nabla F_S(\bm{w}_S^{t})^T \left( \frac{1}{K} \sum_{k=1}^K ( \hat{\nabla} F_{S,k}(\bm{w}_S^{t}) - \nabla F_{S,k}(\bm{w}_S^{t}) ) \right) \right] }_{C_2} \\
&- \underbrace{ \lVert  \mathbb{E} [ \nabla F_S(\bm{w}_S^{t})^T X ] \rVert }_{C_3} \tag{11}
\end{align*}

\noindent $C_1$ is defined as
\begin{align*}
&C_1 = \mathbb{E} [ \lVert \nabla F_{S,k}(\bm{w}_S^{t}) \rVert^2 ] \tag{12}
\end{align*}

\noindent When considering $C_2$
\begin{align*}
&\nabla F_S(\bm{w}_S^{t})^T \left( \frac{1}{K} \sum_{k=1}^K (\hat{\nabla} F_{S,k}(\bm{w}_S^{t}) - \nabla F_{S,k}(\bm{w}_S^{t}) ) \right) \tag{13} \label{eq13} \\
&\geq - \sqrt{G} \frac{1}{K} \sum_{k=1}^K  \left\lVert \hat{\nabla} F_{S,k}(\bm{w}_S^{t}) - \nabla F_{S,k}(\bm{w}_S^{t}) \right\lVert \tag{14} \\
&= - \sqrt{G} \frac{1}{K} \sum_{k=1}^K  \left\lVert \nabla F_{S,k}(\bm{w}_S^{t};\hat{\bm{a}}_k^t) - \nabla F_{S,k}(\bm{w}_S^{t};a_k^t)
\right\lVert \tag{15} \\
&= - \sqrt{G} \frac{1}{K} \sum_{k=1}^K  \bigg|\bigg| \int \nabla \ell ( \hat{a}_k^t;\bm{w}_S) q_{k}^t(a) da \\
&- \int \nabla \ell ( a_k^t;\bm{w}_S) p_{k}^t(a) da
\bigg|\bigg| \tag{16} \\
&\geq - \sqrt{G} \frac{1}{K} \sum_{k=1}^K  \left\lVert \int (\nabla \ell ( \hat{a_k^t};\bm{w}_S) - \nabla \ell ( a_k^t;\bm{w}_S) ) q_{k}^t(a) da \right\lVert \\
&+ \left\lVert \int \nabla \ell ( a_k^t;\bm{w}_S) (q_{k}^t(a) - p_{k}^t(a)) da
\right\lVert \tag{17} \\
&\geq - \sqrt{G} \frac{1}{K} \sum_{k=1}^K  \int \left\lVert \nabla \ell ( \hat{a_k^t};\bm{w}_S) - \nabla \ell ( a_k^t;\bm{w}_S) \right\lVert  \left\lVert q_{k}^t(a)  \right\lVert da \\
&+ \int  \left\lVert \nabla \ell ( a_k^t;\bm{w}_S) \right\lVert  \left\lVert (q_{k}^t(a) - p_{k}^t(a)) 
\right\lVert da \tag{18} \\
&\geq -\sqrt{G} \frac{1}{K} \sum_{k=1}^K (\varepsilon_k^t + \delta_{k}^t) \tag{19} \label{eq19}\\
\end{align*}

\newpage
\thispagestyle{empty}
\noindent Based on Assumption 4 presented in the article, we have
\begin{align*}
& C_2 \geq - \mathbb{E} \left[ \sqrt{G} \frac{1}{K} \sum_{k=1}^K (\varepsilon_k^t + \delta_{k}^t) \right] \tag{20} \\
& \geq -\sqrt{G} \frac{1}{K} \sum_{k=1}^K (H_1+H_2)  \tag{21} \\
& \geq -\sqrt{G} (H_1+H_2)  \tag{22}
\end{align*}

\noindent We consider $C_3$ based on $\mathbb{E} [U^T V] \leq \frac{1}{4} \mathbb{E} [ \lVert U \rVert^2] + \mathbb{E} [ \lVert V \rVert^2], \forall U, \forall V$. Then, we have\begin{align*}
&C_3 = \lVert  \mathbb{E} [ \nabla F_S(\bm{w}_S^{t})^T X ] \rVert \tag{23} \\
&\leq \frac{1}{4}  \mathbb{E} [ \lVert \nabla F_S(\bm{w}_S^{t}) \rVert^2 ] + \mathbb{E} [ \lVert X \rVert^2] \tag{24} \\
&\leq \frac{1}{4}  \mathbb{E} [ \lVert \nabla F_S(\bm{w}_S^{t}) \rVert^2 ] + H_1+H_2 
\tag{25} \label{eq25}
\end{align*}
where Equation~\ref{eq25} can be derived by following similar steps as shown in Equation~\ref{eq13} to Equation~\ref{eq19}.

\noindent By using the results of $C_1$, $C_2$ and $C_3$ we have
\begin{align*}
&B_1 = \mathbb{E} \left[ \nabla F_S(\bm{w}_S^{t})^T \left( \frac{1}{K} \sum_{k=1}^K \nabla F_{S,k}(\bm{w}_S^{t}) \right) \right] \tag{26} \\
& \geq \frac{3}{4} \mathbb{E} [ \lVert \nabla F_{S,k}(\bm{w}_S^{t}) \rVert^2 ]
-(\sqrt{G} + 1) (H_1+H_2) \tag{27} \label{eq27}
\end{align*}

\noindent Now we consider $B_2$
\begin{align*}
&B_2 = \mathbb{E}\left[ \left\lVert \frac{1}{K} \sum_{k=1}^K \nabla F_{S,k}(\bm{w}_S^{t}) \right\rVert ^2 \right] \tag{28} \\
&= \left\lVert \frac{1}{K} \sum_{k=1}^K \nabla F_{S,k}(\bm{w}_S^{t}) \right\rVert ^2 \tag{29} \\
& \underset{a}{\leq} \frac{1}{K} \sum_{k=1}^K \left\lVert \nabla F_{S,k}(\bm{w}_S^{t}) \right\rVert ^2 \tag{30} \\
& \underset{b}{\leq} G \tag{31} \label{eq31}\\
\end{align*}
where $a$ is obtained from the Cauchy-Schwarz inequality and $b$ is obtained from Assumption 2 presented in the article.\\

\noindent By substituting the results of Equation~\ref{eq27} and Equation~\ref{eq31}, we have
\begin{align*}
\mathbb{E}[ F_S(\bm{w}_S^{t+1}) ] 
\leq 
\mathbb{E}[ F_S(\bm{w}_S^{t}) ] 
- \frac{3}{4} \eta_t \mathbb{E} [ \lVert \nabla F_S,k(\bm{w}_S^{t}) \rVert^2 ] \\
+ \eta_t (\sqrt{G} + 1) (H_1+H_2) + \frac{L}{2} \eta_t^2 G 
\tag{32}
\label{eq32}
\end{align*}

\noindent By summing up Equation~\ref{eq32} for all global rounds $t = 0,1, \dots T-1$, we have
\begin{align*}
\mathbb{E}[ F_S(\bm{w}_S^{T}) ] 
\leq 
\mathbb{E}[ F_S(\bm{w}_S^{0}) ] 
- 
\frac{3}{4} \sum_{t=0}^{T-1} \eta_t \mathbb{E} [ \lVert \nabla F_S,k(\bm{w}_S^{t}) \rVert^2 ] \\
+ \sum_{t=0}^{T-1} \left( \eta_t (\sqrt{G} + 1) (H_1+H_2) 
+ \frac{L}{2} \eta_t^2 G \right)
\tag{33}
\end{align*}

\noindent Finally, from $F_S(\bm{w}_S^{*}) \leq \mathbb{E}[ F_S(\bm{w}_S^{T}) ]$ we obtain
\begin{align*}
\frac{1}{\Gamma_T} \sum_{t=0}^{T-1} \eta_t \mathbb{E} [ \lVert \nabla F_{S,k}(\bm{w}_S^{t}) \rVert^2 ] \leq \frac{4 ( F_S(\bm{w}_S^{0}) - F_S(\bm{w}_S^{*}) )}{3 \Gamma_T} \\
+ \frac{1}{\Gamma_T} \sum_{t=0}^{T-1} \left( \eta_t (\sqrt{G} + 1) (H_1+H_2) + 
\frac{L}{2} \eta_t^2 G\right)
\tag{34}
\end{align*}
which completes the proof.

\thispagestyle{empty}
\section{Network, pre-training dataset and partitioning points}
\label{appendix:configuration}
\di{
\subsection{Network bandwidth}
\label{appendix:network}
For all experiments in \mytitle involving different network bandwidth settings, we used real-world data as shown in Table~\ref{table:5g}. We carried out tests on network conditions of 3G HSPA+, 4G LTE-Advanced, and 5G, denoted as 3G, 4G, and 5G, respectively, in this article.
}
\begin{table}[ht]
\begin{center}
\caption{\di{Typical network bandwidths available in the UK.}}
\di{
\begin{tabular}{ P{3.5cm}  P{2cm} P{2cm} }
\Xhline{2\arrayrulewidth}
\textbf{Network type}  &\textbf{Upload bandwidths (Mbps)}  &\textbf{Download bandwidths (Mbps)}\\
\hline
3G&  0.4&  3\\
3G HSPA+&  3&  6\\
4G LTE&   5&  20\\
4G LTE-Advanced&   10&  42\\
Home Broadband Wi-Fi& 11&  60\\
5G&  20&  200\\
\Xhline{2\arrayrulewidth}
\end{tabular}
}
\label{table:5g}
\squeezeup
\end{center}
\end{table}

\di{
\subsection{Pre-training dataset}
\label{appendix:datasets}
}
\begin{table}[ht]
\begin{center}
\caption{\di{Summary of pre-training datasets and training costs. We report the costs for pre-training the VGG11 model as pre-training time in seconds and the proportion relative to the total training time of \mytitle in brackets}}
\di{
\begin{tabular}{ P{1.5cm} P{0.5cm} P{1cm} P{2cm} P{1.8cm}}
\Xhline{2\arrayrulewidth}
Dataset & \#Class  & \#Samples & Visual examples & Pre-training cost\\
\hline
ImageNet  & 1000  & 1.2M &\includegraphics[width=0.9cm]{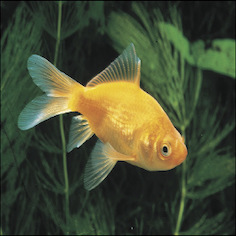}
\includegraphics[width=0.9cm]{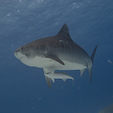}
& n/a\\

Tiny-ImageNet  & 200  & 100K &\includegraphics[width=0.9cm]{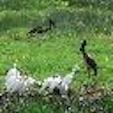} \includegraphics[width=0.9cm]{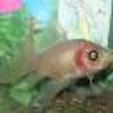}
& 1117s (8\%)\\

CIFAR-5m  & 10 & 50K &\includegraphics[width=0.9cm]{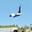}
\includegraphics[width=0.9cm]{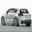}
& 684s (5\%)\\

SIP-17  & 15 & 18K &\includegraphics[width=0.9cm]{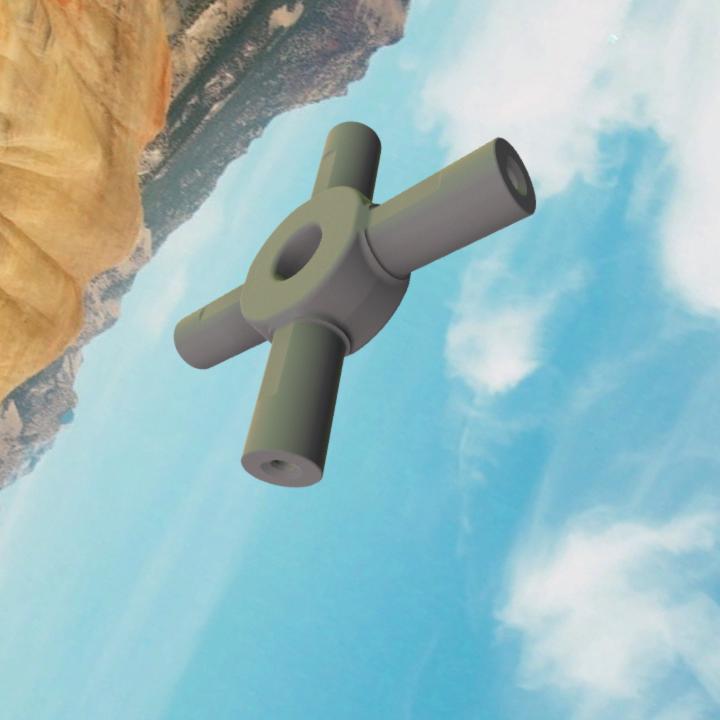}
\includegraphics[width=0.9cm]{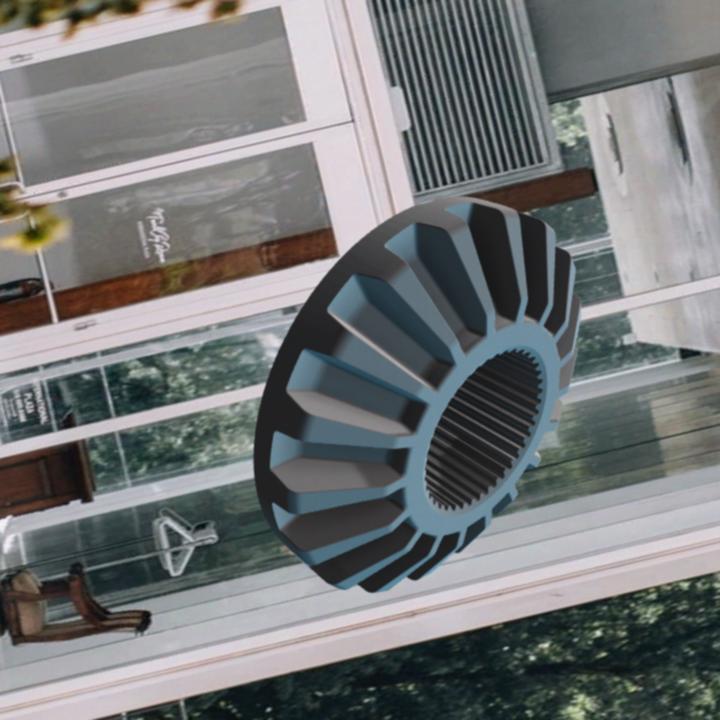}
& 463s (3\%)\\

\Xhline{2\arrayrulewidth}
\end{tabular}
}
\label{table:pre_training dataset}
\end{center}
\end{table}

\di{In this article, we present four different pre-training datasets, namely ImageNet, Tiny-ImageNet, CIFAR-5m, and synthetic industrial parts dataset (SIP-17) shown in Table~\ref{table:pre_training dataset}. 

\textit{ImageNet}~\cite{deng2009imagenet} is a popular large-scale dataset (1.2M samples with 1000 classes) that has been widely used for pre-training on various tasks~\cite{huh2016makes}. 

\textit{Tiny-ImageNet}~\cite{le2015tiny} is a sampled version of ImageNet, containing 100K images from 200 classes, each with 500 training images. 

\textit{CIFAR-5m}~\cite{nakkiran2020deep} is a dataset of 5 million synthetic images. It was generated by sampling from the denoising diffusion probabilistic model (DDPM)~\cite{nakkiran2020deep}. We further sampled the original CIFAR-5m dataset and reduced it to 50K samples, with 5,000 samples per class, to construct a small-scale synthetic dataset that is equal to the size of CIFAR-10. 

\textit{SIP-17}~\cite{zhu2023towards}, consists of 15 individual objects and 2 assembly objects such as crosses and gears, each with 1,200 synthetic images. We used all 15 of the single objects for the pre-training. The SIP-17 dataset is not only synthetic but also exhibits greater domain shift compared to CIFAR-10.

Regarding the pre-training cost, we report the training time of VGG11 model using a single A6000 GPU. For the CIFAR-5m dataset with the same size as CIFAR-10, it took 684 seconds which is equivalent to 5\% of the total training time of \mytitle on our Raspberry Pi prototype.
}

\di{
\subsection{Partitioning point}
\label{appendix:partitioning}
}

\di{
In this article, we employed four different partitioning points (PP) as illustrated in Table~\ref{table:pp}. \mytitle does not optimize partitioning points in DPFL to minimize training latency, but its efficient communication techniques can significantly enhance the performance over a vanilla DPFL system at all partitioning points. In this article, we adopted a fixed partitioning point, i.e., PP2, in the majority of our experiments to align with the configuration of LGL~\cite{hanaccelerating}. However, we also investigate the latency performance of \mytitle for all partitioning points.
}

\di{
\subsection{Accuracy obtained for different partitioning points}
\label{appendix:acc_pps}
}

\begin{table}[ht]
	\centering
	\caption{\di{The highest test accuracy of \mytitle on CIFAR-10 for different partitioning points using  pre-trained ImageNet weights. The results are an average of three independent runs with different random seeds.}}
    \di{
    \begin{tabular}{ P{2.6cm} P{1cm} P{1cm} P{1cm} P{1cm}}
    \Xhline{2\arrayrulewidth}
    \multirow{2}{*}{\textbf{Partitioning points}}&   \multicolumn{2}{c}{\textbf{I.I.D.}}&  \multicolumn{2}{c}{\textbf{Non-I.I.D.}}\\
    \cline{2-5}
     &  \textbf{VGG11}&\textbf{ResNet9}&   \textbf{VGG11}&\textbf{ResNet9}\\
    \hline
    PP1&
    88.36\%&
    88.59\%&
    83.26\%&
    84.34\%\\

    PP2&
    88.87\%&
    88.81\%&
    84.47\%&
    85.82\%\\

    PP3&
    92.6\%&
    92.26\%&
    90.74\%&
    90.42\%\\

    PP4&
    92.34\%&
    91.96\%&
    90.88\%&
    89.44\%\\
    
    \Xhline{2\arrayrulewidth}
    \end{tabular}
    }
    \label{table:acc_pps}
    \squeezeup
\end{table}

\di{
We tested the accuracy of \mytitle for different partition points with pre-trained ImageNet weights on the CIFAR-10 dataset. Overall, \mytitle achieves high accuracy for different partitioning points. It is worth noting that there is a significant accuracy increase with \mytitle when the partition point is moved to later layers (i.e., PP3 and PP4). This indicates that initializing and freezing more layers of pre-trained weights significantly improves the training accuracy of federated learning, which is also observed in prior literature~\cite{chen2022pre}.
}

\newpage
\thispagestyle{empty}
\begin{table*}[ht]
	\centering
	\caption{\di{Different partitioning points (PP) of VGG11 and ResNet9 used for evaluation. Convolution layers denoted as C followed by the no. of filters; filter size is $3 \times 3$ for all convolution layers except for the downsampling convolution where filter size is $1 \times 1$; Max Pooling layer is MP; Fully Connected layer is FC; and Residual Block is RB including two convolution layers, a max pooling layer and downsampling convolution layers; number followed is no. of output channels. We represent communication size as $channels \times width \times height$}}
    \di{
    \begin{tabular}{ P{2cm} P{2cm} P{2cm} P{2cm} P{2cm} P{2cm} P{2cm}}
    \Xhline{2\arrayrulewidth}
    \multirow{2}{*}{\textbf{PP}}&   \multicolumn{2}{c}{\textbf{Device}}& \multicolumn{2}{c}{\textbf{Server}}& 
    \multicolumn{2}{c}{\textbf{Communication size}}\\
    \cline{2-7}
     &  \textbf{VGG11}&\textbf{ResNet9}&   \textbf{VGG11}&\textbf{ResNet9}&
     \textbf{Activation}&\textbf{Gradient}\\
    \hline
    PP1&
    C64-MP&
    C64-MP&
    C128-MP-C256-C256-MP-C512-C512-MP-C512-C512-FC4096-FC4096-FC10&
    C128-MP-RB256-RB512-RB512-FC10&
    $64\times16\times16$&
    $64\times16\times16$\\

    PP2&
    C64-MP-C128-MP&
    C64-MP-C128-MP&
    C256-C256-MP-C512-C512-MP-C512-C512-FC4096-FC4096-FC10&
    RB256-RB512-RB512-FC10&
    $128\times8\times8$&
    $128\times8\times8$\\

    PP3&
    C64-MP-C128-MP-C256-C256-MP&
    C64-MP-C128-MP-RB256&
    C512-C512-MP-C512-C512-FC4096-FC4096-FC10&
    RB512-RB512-FC10&
    $256\times4\times4$&
    $256\times4\times4$\\

    PP4&
    C64-MP-C128-MP-C256-C256-MP-C512-C512-MP&
    C64-MP-C128-MP-RB256-RB512&
    C512-C512-FC4096-FC4096-FC10&
    RB512-FC10&
    $512\times2\times2$&
    $512\times2\times2$\\
    
    \Xhline{2\arrayrulewidth}
    \end{tabular}
    }
    \label{table:pp}
    \squeezeup
\end{table*}